\documentclass[A4]{article} 

\usepackage{ae}
\usepackage{multirow} 
\usepackage{graphicx}
\usepackage{subfigure}
\usepackage{makeidx}
\usepackage{amsmath} 
\usepackage{amssymb}
\usepackage{indentfirst}
\usepackage[hang,small]{caption}

\newcommand{\bs}{\mbox{\boldmath $s$}}
\newcommand{\bc}{\mbox{\boldmath $c$}}
\newcommand{\bb}{\mbox{\boldmath $b$}}
\newcommand{\bu}{\mbox{\boldmath $u$}}

\newcommand{\bx}{\mbox{\boldmath $x$}}

\newcommand{\T}{\scriptscriptstyle\rm T }
\font\bbfnt=msbm10

\def\bbR{\mbox{\bbfnt R}}

\newcommand{\mb}[1]{\mbox{\bfseries \itshape #1}}
\newcommand{\1}{\Bar{1}} 
\newcommand{\2}{\Bar{\Bar{1}}}

\newtheorem{remark}{Remark}[section]
\newtheorem{example}{Example}[section]

\begin{document}

~\vspace{2.0cm}
\begin{center}
\begin{LARGE}
\bf
Structural, Dynamical and Symbolic  \vspace{0.35cm} \\  Observability: From Dynamical
Systems \vspace{0.45cm} \\  to Networks \vspace{1.5cm} 
\end{LARGE}

{\sc  Luis A. Aguirre},    \vspace{0.1cm}

Departamento de Engenharia Eletrônica, Universidade federal de Minas Gerais, 
Av. Antonio Carlos, 6627.  Belo Horizonte, M.G., Brazil. \vspace{0.3cm}

{\sc Leonardo L. Portes}  \vspace{0.1cm}

Programa de Pós-Graduação em Engenharia Elétrica, Universidade federal de Minas Gerais, Av. Antonio Carlos, 6627.
Belo Horizonte, M.G., Brazil.\vspace{0.3cm}

 {\sc Christophe Letellier} \vspace{0.1cm}

Normandie Universit\'e --- CORIA, Campus Universitaire du 
Madrillet, F-76800 Saint-Etienne du Rouvray, France
 \vspace{0.75cm}

{\bf \today}\vspace{1.0cm} 

\end{center}

\section*{Abstract}

The concept of observability of linear systems initiated with Kalman in the mid 
1950s. Roughly a decade later, the observability of nonlinear systems appeared. 
By such definitions a system is either observable or not. Continuous measures 
of observability for linear systems were proposed in the 1970s and two decades 
ago were adapted to deal with nonlinear dynamical systems. Related topics 
developed either independently or as a consequence of these. Observability has 
been recognized as an important feature to study complex networks, but as for 
dynamical systems in the beginning the focus has been on determining conditions 
for a network to be observable. In this relatively new field previous and new 
results on observability merge either producing new terminology or using terms, 
with well established meaning in other fields, to refer to new concepts. 
Motivated by the fact that twenty years have passed since some of these 
concepts were introduced in the field of nonlinear dynamics, in this paper 
(i)~various aspects of observability will be reviewed, and (ii)~it will be 
discussed in which ways networks could be ranked in terms of observability. The 
aim is to make a clear distinction between concepts and to understand what does 
each one contribute to the analysis and monitoring of networks.  Some of the 
main ideas are illustrated with simulations.

\vspace{1.0cm}

\section{Introduction}

One of the many concepts used to analyze dynamical systems and networks is 
observability. The genesis of this can be traced back to mid 20th century. It 
is interesting to see that depending on the research area observability has 
been painted with different colors. In control theory, the cradle of this 
concept \cite{kal/60}, observability is related to the ability of 
reconstructing the state of the system from a limited set of measured variables 
in finite time. A somewhat relaxed version of this definition of observability 
and which is applicable to networks 
is known as structural observability and can be assessed with 
graph \cite{lin/74}.  These 
concepts have a main aspect in common: both classify the system as either being 
observable or not. In this paper the term  {\it structural observability}\, will be used
to refer to such a feature. In the case of networks such concepts could, in principle, 
be used to decide how many nodes should be measured in order to render the 
network observable.

There is a different approach to observability, which evolved from the 
traditional one, that has a different aim. Even if a system is observable, it 
might be advantageous, especially from a numerical point of view, to measure 
specific variables. Instead of a crisp classification in terms of 
observability, this concept permits distinguishing between more and less 
observable scenarios \cite{fri/75,agu_ieee/94}. We shall refer to this as 
{\it dynamical observability}. Two decades ago, some of these concepts were 
adapted to rank variables of nonlinear dynamical systems based on observability 
\cite{let_eal/98} and from there appeared other related approaches that will 
be briefly reviewed in this work.

In the context of networks, the concepts of observability and its dual -- 
controllability -- have been recognized as relevant tools for analysis and 
design \cite{liu_eal/13,wha_eal/15,su_eal/17,lei_eal/17,hab_eal/17}. In this respect, two 
aspects stand out. First, the classical procedures to determine if a system is 
observable face some serious practical and numerical difficulties when applied 
to larger systems.  Indeed, it seems that in the case of 
high-dimensional network, most often the observability is only 
investigated from its topology (described by the adjacency matrix): this will
be referred to as {\it topological observability}\, in this paper. As it will 
be shown, in a network in which oscillators are connected according to an 
adjacency matrix, investigating its connectivity of the corresponding graph is 
not sufficient in general to assess the observability of the network.
Second, to determine a minimum number of sensor nodes for 
which a network is observable is a valuable piece of information. But to be 
able to choose from alternative configurations seems to be an important step 
forward that still has to be accomplished.

As it will be argued in this paper, the classical way of classifying systems as 
being observable or not, cannot really help much in solving the mentioned 
challenge as recently pointed out
\cite{lei_eal/17,hab_eal/17,let_eal/18}. In order to do so, alternative 
scenarios in which observability is guaranteed must be compared in order to 
decide which is more favorable. In other words, as it happened for dynamical 
systems, also for networks there should be a change in paradigm from structural 
to dynamical observability. 

The benefits and need for this has already been pointed out in the literature.
For instance, it has been acknowledged that to choose variables that convey good 
observability of the dynamics enables estimating the state of a network of 
neuron models using Kalman-related methods \cite{sed_eal/12,sch/12}. In a 
recent study about controllability and observability of network topologies built 
with neuron models, it has been found that ``it is necessary to take the node 
dynamics into consideration when selecting the best driver (sensor) node to 
modulate (observe) the whole network activity'' \cite[Sec.\,III-A]{su_eal/17}.
The reader should notice the need to pick {\it good}\, observables or to choose the {\it best}\, sensor nodes.
This type of challenge can be met conceptually using dynamical observability. Of course, the numerical challenge of determining such a property for a large network is of paramount importance and, at the moment, seems unsolved in general.

In view of all this, one of the aims of this paper is to review some concepts 
and procedures concerning observability in the context of nonlinear dynamics. 
It will be useful to see that observability can be classified into different 
types. Hopefully this classification will clarify the main differences which 
could help to answer some of the recent remarks that appeared in the 
literature. Also, the application of such concepts to networks will be 
discussed. Even if from a numerical point of view, some procedures are not 
feasible in the context of large networks, there is much to be gained in 
conceptual terms.  In particular, numerical examples will be provided 
for showing that dynamical observability is not necessarily related
to the network dynamics and that the apparent flow of information can be a
purely dynamical effect.

\subsection{Terminology and organization}

This paper shall refer to dynamical networks as the interconnection of 
dynamical systems of order greater than one. Such dynamical systems will 
sometimes be called oscillators and compose the node dynamics of the dynamical 
network. The interconnection of such nodes is according to a certain topology 
which is described by the adjacency matrix of the network. Graphs can be 
defined for: i)~the node dynamics, which sometimes are referred to as
fluence graphs; ii)~the topology, and for iii)~the full dynamical network 
(combining the node dynamics and the network topology).

This paper is organized as follows.
Section~\ref{ods} reviews a number of concepts that underline the rest of the paper concerning
observability, especially as they emerged from the field of dynamical systems. The counterpart, 
in the context of network topologies, is provided in Section~\ref{gra}. Different types or aspects
of observability are then summarized in Section~\ref{tools}. Section~\ref{net} discusses
the relevance of the aforementioned concepts in the case of nonlinear dynamical networks. 
That section also includes some simulation results. The main points are summarized in 
Section~\ref{conc}, where Table~\ref{tab} is provided as a ``road map'' of this paper.


\section{Observability of Dynamical Systems}
\label{ods}

The objective of this section is 
to give a brief historical background in order to set the remainder of the paper into context. 
The main ideas in this section will be illustrated with the paradigmatic R\"ossler system.

\subsection{Either observable or not}
\label{eon}

The concepts of observability and controllability for linear systems are due to 
Rudolf Kalman \cite{kal/60}. Consider the linear system
\begin{equation}
  \label{abc}
  \left\{
    \begin{array}{lcl}
      \dot{\bx} & = & A \bx + B\bu \\
      \bs & = & C \bx ,
    \end{array}
  \right.
\end{equation}
where $\bx \in \bbR^n$ is the state vector, $\bs \in \bbR^p$ is the measurement 
vector, $\bu \in \bbR^r$ is the input vector and $(A,\,B,\,C)$ are constant 
matrices known respectively as the dynamics matrix, the input or control matrix 
and the output or measurement matrix. The system (\ref{abc}) is said to be 
observable at time $t_f$ if the initial state $\bx(0)$ can be uniquely 
determined from knowledge of a finite time history of the output 
$s(\tau)$, $0 \leq \tau \leq t_f$ \cite{che/99} and the input $u(\tau)$ 
whenever it exists. 

One way of testing whether the system (\ref{abc}) is observable is to define
the {\it observability matrix}:
\begin{eqnarray}
\label{obs1}
 {\cal O} = \left[
  \begin{array}{ccccc}
     C &
     CA &
     CA^2 &
     \ldots &
     CA^{n-1}
  \end{array}
 \right]^{\T} .
\end{eqnarray}

\noindent
The system (\ref{abc}) is therefore observable if matrix ${\cal O}$ is full 
rank, that is if its rank $\rho[{\cal O}]=n$. This is known as Kalman's rank
condition for observability and according to it a pair $[A, C]$ is either
observable or not.

The concepts of controllability and observability were extended to nonlinear systems in the 1970s, e.g.
\cite{her_kre/77}. Consider a nonlinear system
\begin{equation}
\label{b250105} 
 \left\{
    \begin{array}{l}
      \dot{\bx} = \mb{f} (\bx) \\
      s(t) = h (\bx) ,
    \end{array}
  \right.
\end{equation}
with $\mb{f}\,: \bbR^n \rightarrow \bbR^n$ and, for simplicity $s(t) \in \bbR$, that is $h\,: \bbR^n \rightarrow \bbR$. 
Differentiating $s(t)$ yields
\begin{equation}
\label{c250105}
  \dot{s}(t) = \frac{\rm d}{{\rm d}t} h (\bx) = \frac{\partial h}{\partial
  \bx } \dot{\bx}= \frac{\partial h}{\partial
  \bx } \mb{f}(\bx)= {\cal L}_f h (\bx) .
\end{equation}
${\cal L}_f h (\bx)$ is the Lie derivative of $h$ along the vector field 
$\mb{f}$. Hence the time derivatives of $s$ can be written in terms of Lie 
derivatives as $s^{(j)} = {\cal L}^j_f h (\bx)$. The $j$th-order Lie derivative 
is given by 
\begin{equation}
\label{e250105}
  {\cal L}^j_f h (\bx) = \frac{\partial {\cal L}^{j-1}_f h 
  (\bx)}{\partial \bx} \cdot \mb{f} (\bx)
\end{equation}
where ${\cal L}^0_f h (\bx) = h (\bx)$. The observability matrix can be written as
\begin{equation}
\label{newdef}
  {\cal O}_s (\bx) = 
  \left[
    \begin{array}{ccc}
      \displaystyle
      \frac{\partial {\cal L}^0_f h (\bx)}{\partial \bx} &
      \hdots &
      \displaystyle
      \frac{\partial {\cal L}^{n-1}_f h (\bx)}{\partial \bx}
    \end{array}
  \right]^{\T} 
\end{equation}
where the index $s$ has been used to emphasize that ${\cal O}_s
(\bx)$ refers to the system observed from $s(t)$.

The pair $[\mb{f}, h(\bx)]$ in (\ref{b250105}) is said to be {\it observable} 
if $\rho[{\cal O}_s (\bx)]=n,~\forall \bx \in \mathbb{R}^n$,
which is the counterpart of Kalman's rank condition for linear systems -- see \cite{her_kre/77} for details.
If $[\mb{f},\, h(\bx)]$ is observable, any two initial conditions $\bx_{0_1}$ and $\bx_{0_2}$ are
distinguishable with respect to the measured time series $s(t),~t\ge 0$. 
That is, if the system is observable, it is possible to trace back
every single initial condition given only the measured time series
$s(t),~t\ge 0$, or still, $h(\bx_{0_1})|_{t\ge 0} \neq
h(\bx_{0_2})|_{t\ge 0}$ if $\bx_{0_1} \neq \bx_{0_2}$.

Since observability is determined by a rank criterion in both cases,  linear 
and nonlinear systems are classified either as observable or not.

An interesting step in the field was to recognize that the observability matrix in 
(\ref{newdef}) is in fact the Jacobian matrix {\it of the map}\, 
\[ \Phi_s\,: \bbR^n (\mb{x}) \mapsto \bbR^n (s(t),s^{(1)}, ..., s^{(n-1)}),
 \, 
\]
between the original and the $n$-dimensional differential embedding spaces 
\cite{let_eal/05pre}. If  $\Phi_s$ is invertible (injective),  it is possible
to reconstruct the state from $s(t)$. The condition for invertibility of 
$\Phi_s$ at $\mb{x}_0$ is
\begin{equation}
  \label{rank}
  \rho
  \left[
    \left. \displaystyle \frac{\partial \Phi_s}{\partial \mb{x} }
    \right|_{\mb{x} = \mb{x}_0} 
  \right] = n .
\end{equation}
Hence, the system is locally observable if condition (\ref{rank}) holds, that 
is, if $\Phi_s$ is locally invertible. If $\Phi_s$ is constant and invertible, 
then there is a global diffeomorphism and the pair $[\mb{f}, h]$ is fully 
observable. When the reconstructed space is $n$-dimensional, and thus
$\frac{\partial \Phi_s}{\partial \mb{x} }$ is a $n \times n$ matrix, it may be 
also useful to express condition (\ref{rank})
as \cite{let_agu/02}
\begin{equation}
  \mbox{Det }  \frac{\partial \Phi_s}{\partial \mb{x}} \neq 0 \, . 
\end{equation}

\begin{remark}{\rm
If the dimension of the reconstructed space is allowed to increase using
\[ \Phi_s\,: \bbR^n (\mb{x}) \mapsto \bbR^d (s(t),s^{(1)}, ..., s^{(d-1)}) \, ,
\]
with $d>n$, often, singularities that $\Phi_s$ may have will vanish and, then 
$\Phi_s$ gradually becomes full rank. Takens' theorem \cite{tak/81} 
establishes sufficient conditions on $d$ such that $\Phi_s$, for a generic 
measuring function $s(t)=h(\mb{x})$, defines an embedding between the attractor 
in the 
original space and the one in the reconstructed space. Multivariate embeddings 
and some relations between observability theory and Takens' theorem have been 
discussed in \cite{agu_let/05}. Increasing the dimension $d$ in order to remove the 
singularities seems to have serious limitations when networks are considered 
\cite{sen_eal/18}.
}
\end{remark}

\begin{example}{\rm \label{ex1}
Consider the  Rössler system \cite{ros/76} 
\begin{equation}
  \label{roso76}
  \left\{
    \begin{array}{lll}
      \dot{x} & = -y -z \\
      \dot{y} & = x+ay \\
      \dot{z} & = b+z(x-c) , \\
    \end{array}
  \right.
\end{equation}
where $(a,b,c)$ are parameters. If $s=y$, then 
the observability matrix is given by
\begin{equation}
  \label{Qmaty}
  \frac{\partial \Phi_y^3}{\partial \bx} = {\cal O}_{y}(\bx) =
  \left[
    \begin{array}{ccc}
      0 & 1 & 0 \\
      1 & a & 0 \\
      a & a^2-1 & -1
    \end{array}
  \right] ,
\end{equation}

\noindent
where $\Phi_y^3\,: \bbR^3 (\mb{x}) \mapsto \bbR^3 (y(t),y^{(1)},y^{(2)})$
and ${\cal O}_{y}(\bx)$ is constant and nonsingular. Consequently the Rössler 
system is observable from the $y$-variable at any point of the phase space. 
\hfill $\triangle$
\vspace{0.5cm}
}
\end{example}

\subsection{Ranking observable pairs}
\label{ros}

Bernard Friedland suggested computing a conditioning number of a symmetric matrix obtained
from the {\it linear}\, observability or controllability matrices as a way of getting a continuous
function of the parameters instead of a binary (either observable or not) classification \cite{fri/75}.
For the case of observability, Friedland defined the coefficient
\begin{equation}
  \label{do}
  \delta = \frac{\mid \lambda_{\rm min} 
   [{\cal O}^{\T}{\cal O}]\mid}{\mid 
      \lambda_{\rm max} [{\cal O}^{\T}{\cal O}] \mid} ,
\end{equation}
where $\lambda_{\rm max}[{\cal O}^{\T}{\cal O}]$ indicates the maximum 
eigenvalue of  ${\cal O}^{\T}{\cal O}$ (likewise for $\lambda_{\rm min}$). 
Hence even for full row rank observability matrices, the observability 
coefficient $0\le \delta <1$ could be small, indicating ``poor observability''. 
For a nonobservable pair $[A, C]$, $\delta=0$. The following remarks are in 
order. 

\begin{remark}{\rm
The ranking is usually of interest for {\it observable}\, pairs. To make this 
point clear, it will be addressed in the context of single-output linear 
systems, for which $\bc \in \bbR^n$ and the output is given by $\bs=\bc^{\T}\bx$. 
Hence we refer to the observability of the pair $[A, \bc^{\T}]$.
Suppose two pairs $[A, \bc_1^{\T}]$ and  $[A, \bc_2^{\T}]$ have 
observability matrices (see Eq.\,\ref{obs1}) ${\cal O}_1$ and ${\cal O}_2$, 
respectively, such that $\rho[{\cal O}_1]=\rho[{\cal O}_2]=n$, therefore both 
systems are fully observable. Nevertheless, using (\ref{do}) it is found that 
$0<\delta_1<\delta_2$. In such a situation it is said that pair 
$[A, \bc_1^{\T}]$ is less observable than pair $[A, \bc_2^{\T}]$ or, 
alternatively, $s_1 \vartriangleright s_2$ meaning that $s_1=\bc_1^{\T} \bx$ 
(see Eq.\,\ref{abc}) provides better observability of the dynamics in $A$ than 
$s_2=\bc_2^{\T}\bx$. 
}
\end{remark}

\begin{remark}{\rm
A similar result can be stated for nonlinear systems $(\mb{f},\,h_1)$ and $(\mb{f},\,h_2)$. 
}
\end{remark}

\begin{remark}{\rm \label{rem24}
Hence, observability coefficients $\delta$ can be used to rank two pairs with 
$\Phi_{s_1}$ and $\Phi_{s_2}$, which are constant and invertible. This means 
that even if there are global diffeomorphisms, one situation could be 
preferable to the other in a practical setting such as modeling, state 
estimation, experiment and equipment design, and so on. If the
reconstructed space is $n$-dimensional, this can 
be directly assessed by the expression of Det~$\frac{\partial \Phi_s}{\partial 
\mb{x}}$ which can be nonzero but very small in the case of a poor observable.
}
\end{remark}

An example of Remark~\ref{rem24} is provided by the theory of linear systems for which
it is know that similarity transformations of coordinates do not
change the rank of the observability or controllability matrices \cite{che/99}. However, it was shown
that $\delta$ in (\ref{do}) and the counterpart index for controllability {\it are}\, sensitive to similarity 
transformations \cite{agu_ieee/94}, hence
which variable is recorded does matter in practice, even for an observable system.

Following the ideas in \cite{fri/75,agu_ieee/94}, the concept of ranking observable systems
was adapted to nonlinear dynamical systems \cite{let_eal/98,let_agu/02}. In 
particular (\ref{do}) was extended to:
\begin{equation}
  \label{do2}
  \delta_s (\mb{x}) = \frac{\mid \lambda_{\rm min} 
   [{\cal O}_s^T{\cal O}_s,\mb{x}(t)]\mid}{\mid 
      \lambda_{\rm max} [{\cal O}_s^T{\cal O}_s,\mb{x}(t)] \mid} .
\end{equation}
The observability matrix ${\cal O}_s(\bx)$ was originally evaluated using 
(\ref{obs1}) with the Jacobian matrix D$\mb{f}(\bx)$ in place of the dynamics 
matrix $A$. In subsequent works, the observability matrix 
in~Eq.\,(\ref{newdef}) was evaluated along a trajectory $\bx(t),~t_0<t<T$ 
and index (\ref{do2}) averaged along $\bx(t)$, that is
\begin{equation}
\label{do3}
  \delta_s = \frac{1}{T} \int_{0}^{T} \delta_s({\mb{x}(\tau)) {\rm d}\tau},
\end{equation}
where $T$ is the final time considered and $t_0>0$ is chosen to avoid the effect of transients. 
Matrix $[{\cal O}_s^{\T}{\cal O}_s,\mb{x}(t)]$ has been called {\it distortion matrix}\,
in \cite{cas_eal/91}.

\begin{example}{\rm \label{ex2}
For the R\"ossler system (\ref{roso76}), the observability matrix from the $z$ 
variable is
\begin{equation}
  \label{Qmatz}
  \frac{\partial \Phi_z^3}{\partial \bx} =  {\cal O}_{z}(\mb{x}) =
  \left[
    \begin{array}{ccc}
      0 & 0 & 1 \\
      z & 0 & x-c \\
      b+2z(x-c) & -z & (x-c)^2 -y-2z
    \end{array}
  \right],
\end{equation}

\noindent
which is not constant. Besides, because Det(${\cal O}_z)=-z^2$ vanishes for 
$z=0$ this system cannot be ``seen'' from the $z$-variable in the space 
$(z,\,\dot{z},\,\ddot{z})$ when the original system is at 
$\bx=[x,\, y,\, 0]^{\T}$ which is an order-two singular set (the 
so-called singular observability manifold, \cite{fru_eal/12}). Consequently, 
the observability matrix ${\cal O}_z(\mb{x})$ is rank deficient on the singular 
plane $z=0$ and approximately rank deficient  close to that plane.
Using (\ref{Qmaty}), ${\cal O}_x(\bx)$ (not shown) and  (\ref{Qmatz}) in 
(\ref{do2}) and computing (\ref{do3}) the following values were found 
\cite{let_eal/05pre}: $\delta_x=0.022$, $\delta_y=0.133$ and $\delta_z=0.006$, 
hence the variables of the R\"ossler system can be ranked according to 
observability as $y\vartriangleright x \vartriangleright z$, which reflects the
good observability properties conveyed by $y$ and the difficulties of using $z$
to observe the system.
\hfill $\triangle$
\vspace{0.5cm}
}
\end{example}

\subsection{Singularities and lack of observability}
\label{slo}

As illustrated in Example~\ref{ex2}, singularities in the observability matrix indicate 
that the map between the original state space and the considered reconstructed 
space is not globally invertible. In other words, there are ``blind regions'' 
where the system cannot be seen from a particular $d$-dimensional reconstructed 
space. As it will be illustrated in Example~\ref{ex3}, increasing $d$ may eliminate 
singularities in the observability matrix but it should be noted that {\it this is only 
the case for observable systems}. For nonobservable pairs, increasing $d$ 
will not avoid singularities. This can be interpreted as a lack of genericity 
in the measurement function in terms of Takens' theorem.

It will be convenient to distinguish between ``local'' and ``global'' 
singularities. A constant rank-deficient observability matrix will be said to 
have a global singularity because it is always rank-deficient, regardless of 
where the system is in state space. This is always the case for nonobservable 
linear systems. On the other hand, the observability matrix of a 
nonlinear system may become rank-deficient at certain regions of state space. 
For instance, ${\cal O}_{z}(\mb{x})$ in (\ref{Qmatz}) becomes rank-deficient
at $z=0$. The existence of local singularities is a consequence of 
nonlinearity.

A system with a global singularity in its observability matrix is 
nonobservable. This cannot be said of a system with an observability matrix 
with a local singularity. In this case, it is usually more convenient to rank 
the variables based on the singularities that appear in the corresponding 
observability matrices related to the $n$-dimensional space reconstructed using each 
variable. From a practical point of view, the time spent by a trajectory close to a local singularity
will have a direct effect on the observability. This has been used in \cite{fru_eal/12}
to quantify observability.

Hence, observability can be affected by: i)~the choice of 
coordinates of the reconstructed space, and ii)~the existence of 
singularities and the way in which the trajectory relates to them. The first 
case can happen in linear systems as illustrated in \cite{agu_ieee/94} or 
nonlinear systems; the second case only happens in nonlinear systems.

\begin{example}{\rm \label{ex3}
If the Rössler attractor is reconstructed in $(z,z^{(1)},z^{(2)},z^{(3)})$, 
where $z^{(i)}$ is the $i$th derivative of $z$, the corresponding map is 
\cite{agu_let/05} 
\begin{eqnarray}
  \Phi_z^4 =
  \left|
    \begin{array}{l}
      z \\
      \dot{z} = b+z(x-c) \\ 
            \ddot{z} = -(y+z)z + (x-c) [b+ z(x-c)] \\[0.1cm] 
      \stackrel{...}{z} = b(c^2 + x^2) - 2b (cx+y) -(3b+c^3)z \\
      \hspace{1.0cm} + [3(c^2-y-cx)-1+x^2]xz + (3c-a)yz +4(c-x)z^2 \, , 
    \end{array}
  \right. \nonumber
\end{eqnarray}
where the superscript 4 in $\Phi_z^4$ indicates the dimension of the 
reconstructed space. 
Therefore, the observability matrix for $z=0$ (which in Example~\ref{ex2} has been 
shown to be rank deficient on the singular plane in the 3D reconstructed space)
becomes
\begin{eqnarray}
 \left. \frac{\partial \Phi_z^4}{\partial \bx} \right|_{z=0}=
  \left[
    \begin{array}{ccc}x -
      0 & 0 & 1 \\[0.2cm]
      0 & 0 & x-c \\[0.2cm]
            b & 0 & -y + (x-c)^2 \\[0.2cm]
      2b(x\!-\!c) & -2b & -3b -x -ay +(x-c)[-3y +(x-c)^2] 
    \end{array}
  \right], \nonumber
\end{eqnarray}
which is a full column rank matrix. As predicted by Takens' theorem, increasing
the dimension of the reconstructed space removes singularity problems in the 
Jacobian matrix of the coordinate transformation $\Phi_s$. 

This example shows that there is an embedding from $\bbR^3(x,y,z)$ to 
$\bbR^4(z,z^{(1)},z^{(2)},z^{(3)})$ and that the system is observable from such 
a reconstructed space. Alternatively, it can be said that there is a global 
diffeomorphism from the {\it attractor}\, in $\bbR^3(x,y,z)$ to the one in 
$\bbR^4(z,z^{(1)},z^{(2)},z^{(3)})$ -- both attractors have the same dimension. Nonetheless, this was attained at the 
expense of increasing the dimension of the reconstructed {\it space}. This was not 
required for the 
$y$ variable. Hence it is seen that $y$ provides a more favourable situation 
than $z$ and this ranking is quantified by the observability coefficients.  
\hfill $\triangle$
\vspace{0.5cm}
}
\end{example}

\subsection{Graphical approaches}
\label{ga}

Convenient ways of assessing and interpreting observability can be developed 
using graphical techniques. In \cite{let_agu/05pre} a procedure was put 
forward. It consists of representing the variables of {\it a single}\, 
dynamical system and the corresponding relationship by means of a graph that 
resembles an inference diagram. In such a diagram, linear and nonlinear 
dependencies are indicated by continuous and dashed arrows, respectively, as 
shown in the next example.

\begin{example}{\rm \label{ex4}
As a simple example, consider the Rössler system (\ref{roso76}).
The first equation tells us that variables $y$ and $z$ act linearly on $x$. 
Thus, two arrows coming from vertices $y$ and $z$ will reach vertex $x$ with a 
solid line. The second equation can be interpreted likewise. The third equation 
indicates that there is a constant and that variables $x$ and $z$ act nonlinearly 
on $z$. Thus there is a dashed arrow from vertex $x$ to vertex $z$ and 
another one from vertex $z$ to itself. The latter arrow represents the
action of the variable on its own derivatives. When the system is fully 
observable from a variable, the corresponding variable is encircled. The whole 
graph is shown in Fig.\,\ref{gros}a. The solid arrow pointing to $z$ 
represents the constant $b$ in the third equation.

\begin{figure}[ht]
  \begin{center}
    \includegraphics[scale=0.85]{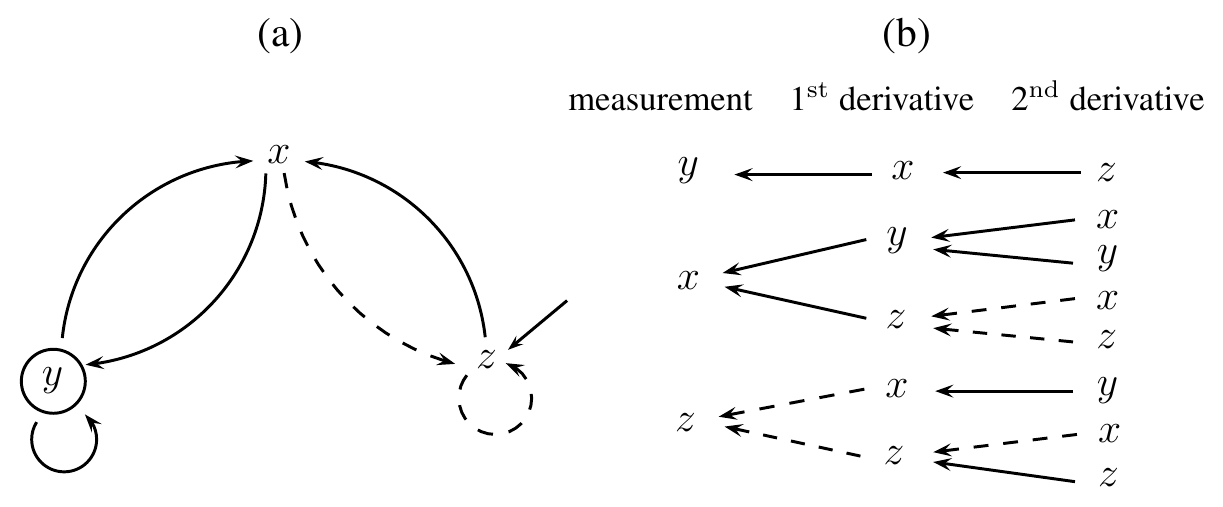} \\[-0.3cm]
    \caption{(a)~Graph of the interaction between the variables for the
Rössler system. A solid (dashed) arrow represents a (non) linear coupling. (b)~Unfolded 
schematic view of the variables reached when the first 
and the second derivative are computed. For the sake of observability, it
is more detrimental when a dashed arrow connects the measured variable to the first
derivative, than when it connects the first derivative to the second. Hence, $x$ conveys
better observability than~$z$.}
    \label{gros}
  \end{center}
\end{figure}

From the graph in Figure~\ref{gros}a, an unfolded scheme is built by graphically
visiting the vertices starting from the measured variable $s(t)$, 
and moving against the arrow directions one moves one step for each
additional dimension in the reconstructed space. 

Figure~\ref{gros}b shows the case for a 3D differential reconstructed space 
$(s(t),s^{(1)},s^{(2)})$, where $s=h(\bx)$ is the recorded variable. Whenever 
the three variables $(x,\,y,\,z)$ are connected horizontally by solid arrows
there is a global diffeomorphism and the observability is complete. Hence in
Figure~\ref{gros}b it is seen that only for $s=y$ there is a global diffeomorphism
between the original state space and $(y,\,\dot{y},\,\ddot{y})$. At the other
extreme, notice that there are dashed lines in the first stage of this diagram 
-- moving from left to right -- 
when the measured variable is $z$. Measuring variable $x$ is somewhere in between, hence
$y\vartriangleright x \vartriangleright z$.
\hfill $\triangle$
\vspace{0.5cm}
}
\end{example}

Although this procedure does not result in numerical indices, it falls into the
category of ranking observable systems. This is an important point because, as it
will be seen later in Sec.\,\ref{gra}, there are other graphical procedures that follow the 
{\it either observable or not}\, framework.

\subsection{Symbolic Observability}
\label{so}

As discussed in Sec.\,\ref{slo}, one of the aspects that greatly influence 
observability in nonlinear systems are the singularities that appear in the 
observability matrix. Because at a singularity the determinant of the $n \times 
n$ observability matrix will become null, the underlying motivation in symbolic 
observability is that the more complicated the determinant 
Det[$\mathcal{\tilde{O}}_s$] of the symbolic observability matrix, the less 
observable the system is. A first approach to symbolic observability was described 
in \cite{let_agu/09pre}, where only polynomial elements were used. 

However the analytical computation of Det[$\mathcal{\tilde{O}}_s$] can be
a nearly impossible task for a five-dimensional rational system. Nevertheless
the complexity of Det[$\mathcal{\tilde{O}}_s$] can be assessed simply by counting the number of 
linear, nonlinear and rational terms in it, without paying attention
to its exact form and this will suffice to quantify observability \cite {bia_eal/15}. 

The main steps for computing symbolic observability indices are: i)~obtain the symbolic
Jacobian matrix $\mathcal{\tilde{J}}$ from the classical Jacobian matrix by replacing  
linear, nonlinear and rational elements, respectively with 1, $\1$, and $\2$; ii)~build the symbolic 
observability matrix $\mathcal{\tilde{O}}_s$ as detailed in \cite{bia_eal/15}, 
iii)~compute the symbolic expression for
Det[$\mathcal{\tilde{O}}_s$] and count the number of symbolic terms in such an expression,
iv)~finally, the symbolic observability coefficient is obtained as
\begin{equation}
\label{soc}
  \begin{array}{rl}
    \eta_{s^n} =  & \displaystyle
    \frac{N_1}{N_1 + N_{\1} + N_{\2}}     + \frac{N_{\1}}{\left( \displaystyle \mbox{max}(N_1,1) + N_{\1} 
          + N_{\2} \right)^2} \\[0.4cm]
    & \displaystyle
    + \frac{N_{\2}}{\left( \displaystyle \mbox{max}(N_1,1) + N_{\1} 
          + N_{\2} \right)^3} ,
  \end{array}
\end{equation}

\noindent
where $N_1$, $N_{\1}$ and $N_{\2}$ are the numbers of symbolic terms 1, $\1$ 
and $\2$, respectively.

\begin{example}{\rm \label{ex25}
For the R\"ossler system (\ref{roso76}), the Jacobian and symbolic Jacobian 
matrices are
\begin{equation}
  \label{JJ}
  {\rm D}\mb{f} =  
  \left[
    \begin{array}{ccc}
      0 & -1 & -1 \\
      1 & a & 0 \\
      z & 0 & (x-c)
    \end{array}
  \right] \mbox{ and }
   {\cal \tilde{J}} =  
  \left[
    \begin{array}{ccc}
      0 & 1 & 1 \\
      1 & 1 & 0 \\
      \1 & 0 & \1
    \end{array}
  \right]  \, , 
\end{equation}
respectively.  
Notice that ${\cal \tilde{J}}$ can be obtained from ${\rm D}\mb{f}$\, by inspection. If variable $x$ is measured,
the respective observability matrix is given by \cite{bia_eal/15}:
\begin{equation}
  \label{Ox3}
  {\cal \tilde{O}}_x =  
  \left[
    \begin{array}{ccc}
      1 & 0 & 0 \\
      0 & 1 & 1 \\
      \1 & 1 & \1
    \end{array}
  \right],
\end{equation}

\noindent
for which the symbolic determinant is Det[$\mathcal{\tilde{O}}_x]=1\otimes(1\otimes \1 -1\otimes 1)$.
In that expression there are four $1$s, and one $\1$, hence $N_1=4$, $N_{\1}=1$ and $N_{\2}=0$. Using these
values in (\ref{soc}) yields $\eta_{x^3}=0.84$. In a similar way the other symbolic coefficients can be
readily obtained \cite{bia_eal/15}: $\eta_{y^3}=1$ and $\eta_{z^3}=0.56$, where the exponent indicates
the dimension of the reconstruction space (see Example~\ref{ex3}). 
Therefore the variables can be ranked as before $y\vartriangleright x \vartriangleright z$.
\hfill $\triangle$
\vspace{0.5cm}
}
\end{example}

All the types of observability discussed so far are defined based on the system 
equations. In experimental situations, these equations are rarely known. An 
indirect way of accessing observability from data will be briefly mentioned in what 
follows.

\subsection{Data-based Observability}
\label{dbo}

Motivated by the fact that in practice the system equations are not always 
available, an alternative procedure for assessing observability was proposed in 
\cite{agu_let/11}. However, observability is, by definition, related to the
equations of the vector field or related to the map, in the case of 
discrete-time systems. Hence estimating coefficients from data is only an 
indirect way of assessing observability from some of its {\it signatures}\, 
found in a reconstructed space, as explained next.

The rationale behind the method in \cite{agu_let/11} is that in the 
reconstructed space of a system with poor observability conveyed by a recorded 
time series, trajectories are either pleated or squeezed. Such features result 
in a more complex {\it local}\, structure in the reconstructed space. On the 
other hand, in the space reconstructed using good observables, very often, 
trajectories are unfolded comfortably and that translates into a more simple 
local structure of such a space. The SVDO (singular value decomposition 
observability) coefficients hence quantify, using the singular value 
decomposition (SVD) of a trajectory matrix, the local complexity of the 
reconstructed space. Simpler structures are associated to better observability 
whereas more complex local structures are related to poorer observability. 

A key point to be noticed here is that SVDO cannot quantify observability 
{\it per se}, which by definition
would require the vector field equations, but rather are indicators of the {\it average local complexity}\,
of a reconstructed space, which often -- but not always -- correlates with observability. This remark
seems to be general. Therefore by {\it data-based observability}\, we refer to the {\it indirect}\, quantification
of observability without the use of the system equations.

\section{Graphical Approaches for Assessing Observability }
\label{gra}

This section is devoted to graph-theoretic approaches for 
assessing observability of dynamical systems. When a network is considered, 
there are three levels of description: i) the node dynamics, commonly made of a
dynamical system (oscillator), ii) the topology of the network, described by 
the corresponding adjacency matrix, and iii) the full network combining the 
node dynamics with the network topology. Each level can be represented by a 
specific graph providing different assessment of the network observability 
as it will be addressed in Sec.\,\ref{net}. Given the importance of graphs, 
this section reviews some results concerning the quantification of 
observability from such a representation. Some examples will be taken using 
simple dynamical systems (oscillators).

\subsection{Lin's method}
\label{lm}

In a seminal paper, Lin developed the concept of structural controllability 
\cite{lin/74} which was later extended to that of structural observability in 
\cite{cha_sha/92}. Such concepts have been defined for linear systems as 
(\ref{abc}). In words, a linear dynamical pair $[A, C]$ is structurally 
observable if there exists a ``perturbed'' pair $[A_1, C_1]$ of the same 
dimension with the same structure which is completely observable. $[A, C]$ and 
$[A_1, C_1]$ are of the same structure if for every fixed zero entry of 
$[A, C]$ the corresponding entry of the pair $[A_1, C_1]$ is also a fixed 
zero and vice-versa \cite{cha_sha/92}. Also, $[A_1, C_1]$ is a perturbed pair 
of $[A, C]$ in the sense that there exists an $\epsilon>0$ such that 
$||A-A_1||<\epsilon$ and $\parallel C-C_1\parallel<\epsilon$. For instance, 
consider the pair
\begin{eqnarray}
\label{nonobs}
A=\left[
\begin{array}{cc}
A_{11} & 0 \\
A_{21} & A_{22}
\end{array}
\right],~ 
C=\left[
\begin{array}{cc}
C_1 & 0
\end{array}
\right], 
\end{eqnarray}

\noindent
where the nonzero entries can assume any values. Clearly, the observability 
matrix (\ref{obs1}) will be rank deficient regardless of the values of $A_{ij}$ 
and of $C_1$. Notice that for each values given to $A_{ij}$ and of $C_1$ the 
resulting pair will be of the same structure, and still nonobservable. Hence
the pair (\ref{nonobs}) is (structurally) nonobservable. This concept of 
structural observability (the same applies to controllability) is closely 
related to the determination of observability by inspection from a system 
represented in Jordan canonical form, where what matters is the location of 
zero and non zero elements in the pair $[A, C]$ \cite{che/99}.

A very interesting analysis proposed by Lin was the drawing of a graph for the pair 
$[A, \bb]$. Suppose the said pair is
\begin{eqnarray}
A=\left[
\begin{array}{ccc}
a_{11} & a_{12} & 0 \\
a_{21} & a_{22} & 0 \\
a_{31} & a_{32} & a_{33}
\end{array}
\right],~ 
\bb=\left[
\begin{array}{c}
0\\
0\\
b_3
\end{array}
\right], \nonumber
\end{eqnarray} 

\noindent
in which the only fixed values are the zeros and the remaining entries can take 
any values. In order to build the graph, pair $[A,\bb]$ is rewritten as
\begin{eqnarray}
\label{b060917}
& 
\begin{array}{ccccc}
v_1~ & v_2~ & v_3~ & \vdots & v_4 
\end{array} \nonumber \\
& [A \, \vdots \, \bb]= \left[
\begin{array}{ccccc}
a_{11} & a_{12} & 0 & \vdots & 0\\
a_{21} & a_{22} & 0 & \vdots & 0\\
a_{31} & a_{32} & a_{33} & \vdots & b_3
\end{array}
\right]~
\begin{array}{c}
\rightarrow v_1 \\
\rightarrow v_2 \\
\rightarrow v_3 .
\end{array} \nonumber
\end{eqnarray} 

\begin{figure}[ht]
  \begin{center}
    \includegraphics[scale=0.7]{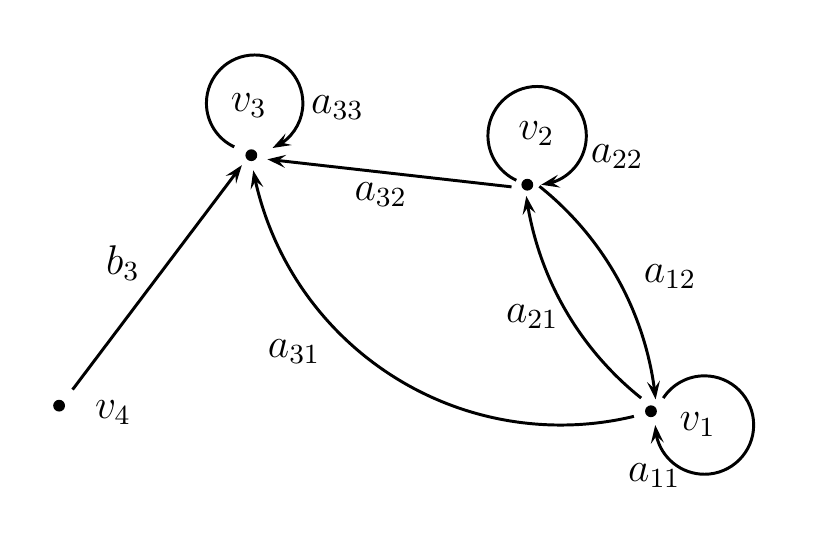} \\[-0.2cm]
    \caption{\label{f060917a}Graph of pair $[A, \bb]$. In this graph 
vertices $v_1$ and $v_2$ are {\it non-accessible}\, from the input vertex $v_4$ 
and, consequently, the pair is non structurally controllable.}
  \end{center}
\end{figure}

\noindent
Each column of $[A\, \vdots \,\bb]$ labeled $\{v_1,\, v_2,\, v_3,\, v_4\}$ defines a 
vertex, (see Fig.\,\ref{f060917a}). In the sequel, the oriented edges must be 
determined. To this end, for each origin vertex (column), the location of the 
nonzero elements indicate to which vertex the edge points to.  Hence the nonzero 
element in the fourth column, row three (element $b_3$) indicates that there is 
an oriented edge from $v_4$ to $v_3$. This must be done for all the columns of 
$[A\, \vdots \, \bb]$. The result is shown in Figure~\ref{f060917a}, where the values of 
the entries in $[A\, \vdots \,\bb]$ are shown as weights. In the sequel, Lin showed that 
the graph of an uncontrollable pair $[A,\bb]$ has {\it non-accessible}\, nodes 
as in Figure~\ref{f060917a}. Hence the contribution of his work is to define 
structural controllability in terms of necessary (but not sufficient) graph 
properties.

An extension of Lin's procedure to build a graph and determine if it is 
structurally controllable for the case of observability can be easily 
accomplished by means of the {\it duality theorem}\, \cite{che/99}. Hence, the 
pair $[A, \bc^{\T}]$ (see Remark~2.2) is structurally observable iff its dual $[A^{\T},\,\bc]$ is 
structurally controllable. When matrix $A$ is transposed, the arrows of the 
edges should point in the reverse direction.

\begin{example}{\rm \label{ex5}
In this example it is shown how the R\"ossler system (\ref{roso76}) can be 
represented using a graph such that a procedure akin to Lin's can be followed. 
Notice that Lin's results were developed neither for observability nor for 
dynamical networks, but rather he showed how to represent {\it a single linear 
system}\, as a graph and then described Kalman's rank condition in terms of 
graph properties. Hence his starting point is the dynamic matrix $A$ and the 
input vector $\bb$. The controllability of the R\"ossler system can be
investigated using the Jacobian matrix D$\mb{f}$ of  
(\ref{roso76}) and an input vector $\bb$:
\begin{eqnarray}
  \label{a070917b}
& 
\begin{array}{ccccccc}
\hspace{0.6cm} x & ~ y~ & ~ ~z &  \hspace{0.2cm}b_x& \!\!\vdots  & \!b_y \!& \!b_z 
\end{array} \nonumber \\
& [{\rm D}\mb{f} ~\vdots \, \bb] = 
  \left[
    \begin{array}{ccccccc}
       0 & -1 & -1 & 1 & \vdots & 0 & 0\\
       1 & a & 0 & 0 & \vdots & 1 & 0\\
       z & 0 & x-c & 0 & \vdots & 0 & 1\\
    \end{array}
  \right]~
  \begin{array}{c}
    \rightarrow x \\[0.2cm]
    \rightarrow y \\[0.2cm]
    \rightarrow z 
  \end{array} \nonumber
\end{eqnarray} 

\noindent
Figure~\ref{f070917a}a shows the graph of pair 
$[{\rm D}\mb{f}, [0\,0\,1]^{\T}]$.  Vertices $x$ and $y$ are both accessible 
from vertex $b_z$: the R\"ossler system is structurally controllable when the 
system is driven from the $b_z$ vertex. When the control is 
applied to variable $y$, vertex $x$ is accessible but vertex $z$ will not 
be accessible if the dashed link vanishes ($z=0$): the pair $[{\rm D}\mb{f}, [0\,1\,0]^{\T}]$ is therefore not 
structurally controllable for $z=0$. A similar result is obtained for 
the pair $[{\rm D}\mb{f}, [1\,0\,0]^{\T}]$.

\begin{figure}[ht]
  \begin{center}
    \begin{tabular}{ccc}
      \includegraphics[width=0.26\textwidth]{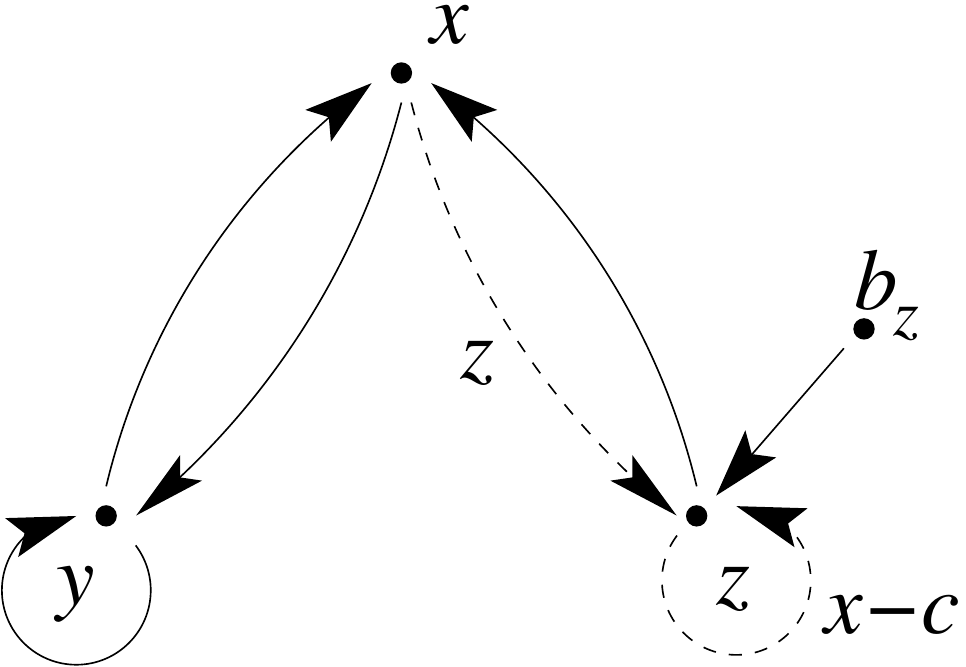} &
      \includegraphics[width=0.26\textwidth]{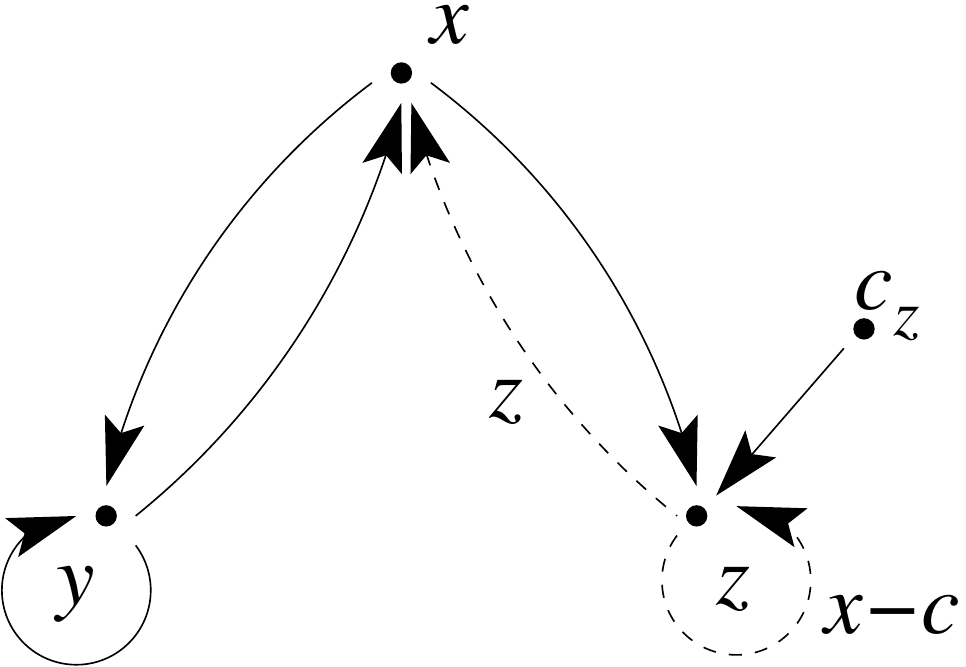} &
      \includegraphics[width=0.26\textwidth]{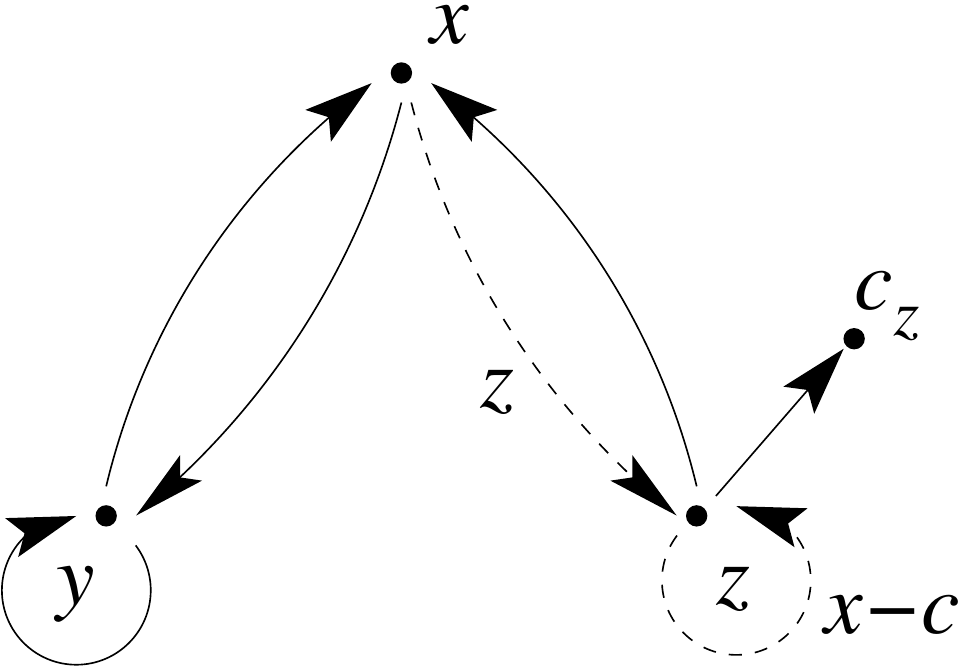} \\     
            {\small (a) Graph of $[{\rm D}\mb{f},\bb_z]$} & 
      {\small (b) Graph of $[({\rm D}\mb{f})^{\T},\bc_z]$}  &
      {\small (c) Graph of $[{\rm D}\mb{f},\bc_z^{\T}]$} \\[-0.1cm]
    \end{tabular}
    \caption{(a)~Graph of the pair $[{\rm D}\mb{f}, \bb_z]$ where 
$\bb_z=[0~0~1]^{\T}$ is the input vector. (b)~Graph of 
$[({\rm D}\mb{f})^{\T},\bc_z]$, the dual of (a), where $\bc_z^{\T}=[0\,0\,1]$ is the 
input vector. (c)~Graph of $[{\rm D}\mb{f}, \bc_z]$ used with the ``dual 
interpretation''. Dashed lines indicate non constant connections due to 
nonlinearities. Notice the similarity with the graph in Figure~\ref{gros}a. }
    \label{f070917a}
  \end{center}
\end{figure}

In order to investigate the observability using Lin's result, we have to use 
the transposed Jacobian matrix (D$\mb{f})^{\T}$ of (\ref{roso76}) and the 
vector $\bc$. This is called the dual system of $[{\rm D}\mb{f}, \bb] $
when $\bc=\bb$. The graph in Figure~\ref{f070917a}b,  when $z$ is measured, 
$\bc_z^{\T} = [0\,0\,1]$ can be obtained as before using: 
\begin{eqnarray}
\label{a070917}
& 
  \begin{array}{cccccc}
    \hspace{1.0cm} ~x & ~ ~y~ &  ~z &  \hspace{0.3cm} \vdots & c_z   
  \end{array} \nonumber \\
    & [({\rm D}\mb{f})^{\T} ~\vdots \,\bc_z]= 
  \left[
    \begin{array}{cccccccc}
      0 & 1 & z & \vdots & 0 \\
      -1 & a & 0 & \vdots & 0 \\
      -1 & 0 & x-c & \vdots & 1 \\
    \end{array}
  \right]~
  \begin{array}{c}
    \rightarrow x \\[0.18cm]
    \rightarrow y \\[0.18cm]
    \rightarrow z 
  \end{array} \nonumber
\end{eqnarray} 

\noindent
It should be noticed that at $z=0$ the connection from vertex $z$ to vertex $x$ 
vanishes and both $x$ and $y$ become non-accessible vertices 
(Fig.~\ref{f070917a}b). Hence at $z=0$ the pair 
$[({\rm D}\mb{f})^{\T},\bc]$ is noncontrollable. From the duality 
theorem, this implies that the pair $[{\rm D}\mb{f},\bc^{\T}]$ is not observable 
at $z=0$, as seen before in Example~\ref{ex2}. 

We can reach a similar conclusion from the graph shown in Fig.\ \ref{f070917a}a 
but drawing an output vector $\bc$ (Fig.\ \ref{f070917a}c) and using a ``dual 
interpretation'' for the edges. Thus, an edge from vertex $v_i$ to vertex 
$v_j$ means that $v_j$ receives information from $v_i$. Figure~\ref{f070917a}c 
illustrates the case when $z$ is measured, hence an {\it output edge}\, $c_z$ 
is drawn. Because the flow of information from $x$ -- and consequently from $y$ 
-- is cut when $z=0$, the pair $[{\rm D}\mb{f},\,[0\,0\,1]]$ is 
structurally nonobservable. If we proceed in this way, it is found that the 
pair $[{\rm D}\mb{f},\,[0\,1\,0]]$ is structurally observable. This is in 
agreement with the fact there exists a global diffeomorphism between the 
original state space and $(y, \dot{y}, \ddot{y})$ \cite{let_eal/05pre}. 
\hfill $\triangle$
\vspace{0.5cm}
}
\end{example}

From the discussion above, it is clear that structural observability is unable 
to distinguish, given an {\it observable}\, system, situations with different 
observability features. For instance, for $0<z\ll 1$ the edge linking $z$ to 
$x$ in Figure~\ref{f070917a} has not yet vanished and the system remains 
structurally observable as well as for another system for which such a link has 
a constant weight. Hence this way of addressing the observability of a graph
is overcome by other definitions of observability.

It is important to notice that as a consequence of nonlinearity there will be 
non constant elements in the matrix $[({\rm D}\mb{f})^{\T} \, \vdots \,\bc]$ and 
therefore there will be dashed connections (that can vanish) in the graph. 
Hence procedures to investigate observability that treat constant and 
variable connections alike ignore the effect of nonlinearity which is one of the 
main causes of singularities which, in turn, greatly affect the observability 
of a system, as discussed in Sec.\,\ref{slo}.

\subsection{Liu and coworkers' method: sensor sets}
\label{lcm}

A more recent procedure has been put forward by Liu and coworkers who have 
addressed the problem of
determining the minimum number of sensor nodes needed to reconstruct the state
\cite{liu_eal/13}. The suggested graphical approach is claimed to provide
a necessary and sufficient sensor set to render the system observable. 

First, an inference diagram is built, this is a graph. The graph is decomposed 
in strongly connected components (SCC) which are the largest subgraphs in which 
there is a directed path from every vertex to any other vertex. If an SCC does 
not have any incoming edges, it has been called a root SCC \cite{liu_eal/13}. 
Observability of the whole system is said to be achieved if at least one vertex 
of each root SCC is measured.

\begin{figure}[ht]
  \begin{center}
    \begin{tabular}{ccc}
      \includegraphics[scale=0.34]{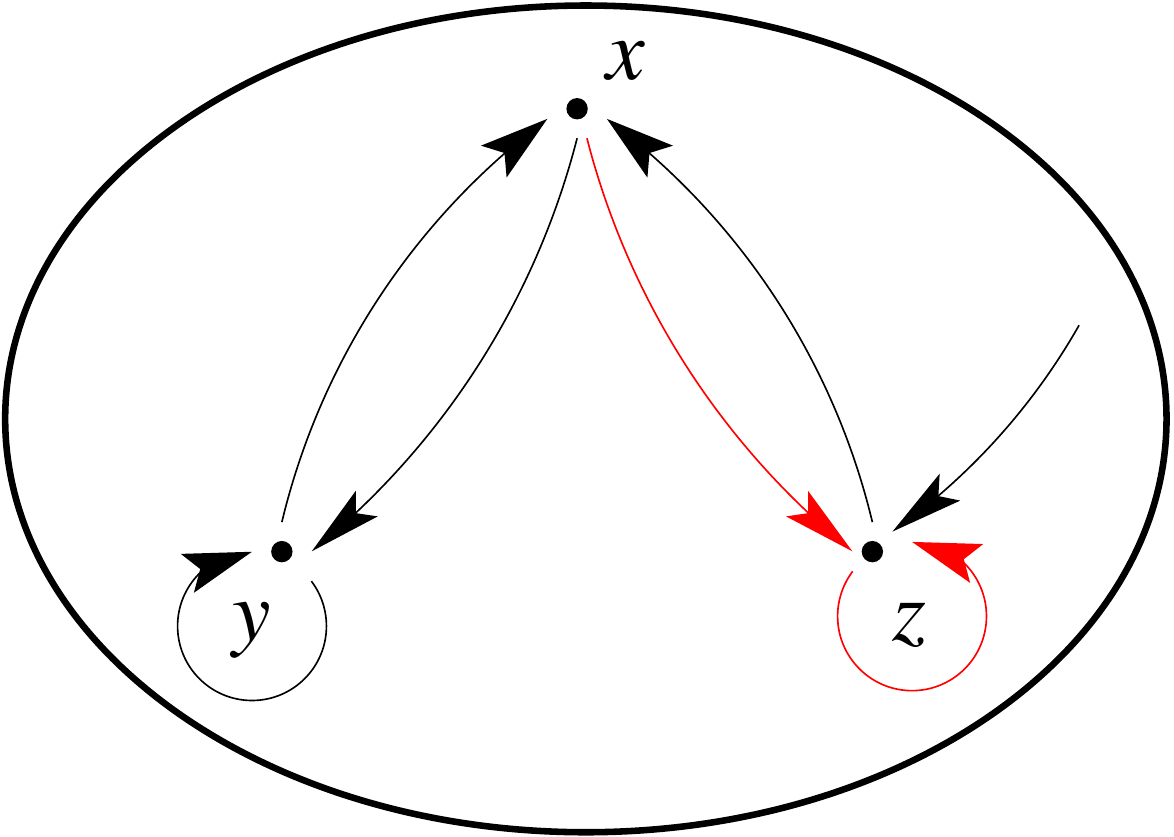} & ~~~~ &
      \includegraphics[scale=0.34]{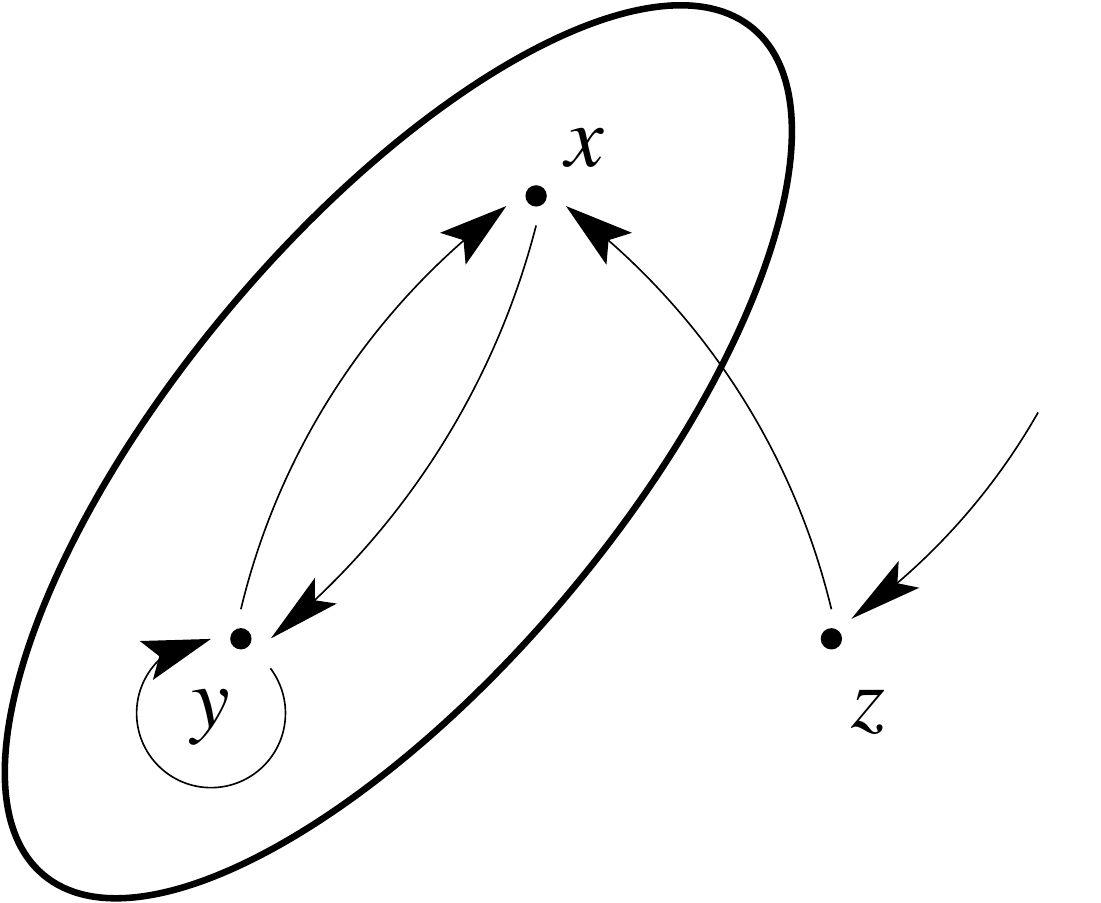} \\
      {\small (a) Graph with all links} &  ~~~~ &
      {\small (b) Graph with only linear links} \\[-0.2cm]
    \end{tabular}
    \caption{\label{GrafosXsymbJ}Graphs of the R\"ossler system.
The root SCC (drawn as a thick circle) contains the three variables when the
method considers all connections constant (a) and only variables $x$ and $y$ 
when nonlinearities are removed for building the graph (b).}
  \end{center}
\end{figure}

\begin{example}{\rm \label{ex32}
We start with the graph shown in Figure~\ref{gros}a which corresponds to the 
R\"ossler system (\ref{roso76}) but without distinguishing between full and 
dashed lines. Notice that it is possible to start at any vertex (node or 
variable) and reach all other vertices following the arrows. Hence, the whole 
graph is an SCC. Because there is no incoming edge, this is also a root SCC. 
Hence in order to guarantee observability it suffices to measure any of its 
variables. However, if the dashed line vanishes, the $z$ variable will no longer
be part of the SCC  (see Figure~\ref{GrafosXsymbJ}b) and should not be measured.
\hfill $\triangle$
\vspace{0.5cm}
}
\end{example}

Example~\ref{ex32} shows that this method, as acknowledged by the authors 
\cite[p.\,2464]{liu_eal/13} is unable to indicate that measuring the $y$ 
variable from the R\"ossler system is preferrable to, say, measuring $z$. 
On the other hand, it was shown that this graphical approach 
underestimates the number of variables which must be necessarily measured
\cite{hab_eal/17,let_eal/18}. An  improved version of this graphical approach 
was recently proposed \cite{let_eal/18b},  showing that nonlinear interactions
should be removed for determining the root SCCs and that such graph only 
provides necessary but not sufficient conditions on the measurements for
ensuring structural observability.

\subsection{Ranking observable graphs}
\label{rog}

Lin's method for structural observability in Sec.\,\ref{lm} was developed for 
linear systems for which the weights in the corresponding graph are constant. 
In determining non-accessible vertices only the presence or absence of edges is 
of concern. Therefore the method either classifies the graph as observable or 
not.

A more challenging situation is furnished by the pair $[A,\,\tilde{C}]$ with 
$A$ given in (\ref{nonobs}) and $\tilde{C}=[C_1~C_2]$, as follows. If 
$C_2\neq 0$ the pair stands a chance of being observable. Let us assume that it 
is observable, that is, the observability matrix (\ref{obs1}) computed with the 
pair $[A,\tilde{C}]$ is full rank, hence the pair is structurally observable 
for the reasons given above. Structural observability will be lost only if 
$C_2= 0$, and even for extremely small values of $C_2$, the pair will be 
structurally observable. Hence such type of observability will not distinguish 
among a whole range of pairs that can be either far or arbitrarily close to the 
condition $C_2= 0$. A possible way out in this very simple example is to 
compute the condition number (\ref{do}) for the observability matrices of 
$[A, \tilde{C}]$ for the different measuring situations that result in 
different $\tilde{C}$s. Ill-conditioned observability matrices will indicate 
unfavorable situations in terms of observability.

As for the method by Liu and coworkers for sensor set selection, the lack of 
discriminatory power pointed out in Example~\ref{ex32} is due to disregarding 
the differences in the type of edges, that is, the method treats full and 
dashed arrows alike. In order to rank the variables, features of the links 
should be taken into account, such as the weight of a link: small constant 
weights and variable weights will give rise to poorly observed regions in the 
graph.

It is interesting to notice that as it happened in the development of the 
theory of observability for dynamical systems, the first results classified 
graphs either as being observable or not. It seems that it would be desirable 
to see the development of procedures to rank graphs in terms of observability.

\subsection{Symbolic observability of topologies}
\label{sog}

Provided that the symbolic Jacobian matrix ${\cal J}$ can be written for a 
graph then, in principle, symbolic observability coefficients can be computed. For 
relatively simple systems, to obtain ${\cal J}$ is straightforward, as the following 
example shows.

\begin{example}{\rm \label{ex33}
We again consider the graph shown in Figure~\ref{gros}a.  In a typical graph, 
there would be no distinction between full and dashed lines, as for the methods 
of Lin and of Liu and coworkers. Calling ${\cal J}^0$ a symbolic Jacobian 
matrix that does {\it not}\, take into account the nonlinear connections, and 
${\cal \tilde{J}}$ the standard symbolic Jacobian matrix \cite{bia_eal/15}, 
from system (\ref{roso76}) we get
\begin{equation}
  \label{JJ2}
  {\cal J}^0 =  
  \left[
    \begin{array}{ccc}
      0 & 1 & 1 \\
      1 & 1 & 0 \\
      1 & 0 & 1
    \end{array}
  \right] \mbox{ and }
   {\cal \tilde{J}} =  
  \left[
    \begin{array}{ccc}
      0 & 1 & 1 \\
      1 & 1 & 0 \\
      \1 & 0 & \1
    \end{array}
  \right]  .
\end{equation}

\noindent
Notice that ${\cal \tilde{J}}$ is the same as obtained in (\ref{JJ}). Hence 
proceeding as in Example~\ref{ex25} the same symbolic observability 
coefficients obtained from the system equations are found using 
${\cal \tilde{J}}$, that is, from the graph. If ${\cal J}^0$ is used instead, 
the result reached at is that any of the variables provide the same level of 
observability. This shows why the method by Liu and co-workers is unable to 
provide guidance of which sensor vertex to use within the root SCC which 
here (Figure~\ref{GrafosXsymbJ}a) contains the three variables. In the spirit
of symbolic coefficients, the modified approach \cite{let_eal/18b} does not
take into account the nonlinear (dashed) edges (Figure~\ref{GrafosXsymbJ}b).
\hfill $\triangle$
\vspace{0.5cm}
}
\end{example}

It is conceivable that for graphs of even moderate sizes, it might not be 
simple to build analytical observability matrix
and even less to compute the determinant of the symbolic observability matrix. 
A software like Maple fails to compute the observability matrix of a 5D 
rational system \cite{let_eal/18}.  A similar difficulty is shared by all 
other methods that require the analysis of an observability matrix. 
Symbolic approaches are therefore an alternative to overcome
this difficulty.

\section{Types of Observability}
\label{tools}

The aim of this section is to recognize differences among types of 
observability in what concerns definitions and aims, as reviewed in 
sections~\ref{ods} and~\ref{gra}. Also, interesting links between definitions 
will be pointed out and some extensions to networks will be proposed. The types 
of observability will be mentioned roughly in the same chronological order as 
they appear in the literature. The main results are summarized in 
Table~\ref{tab}.

\subsection{Structural Observability}
\label{sto}

The adjective {\it structural}\, was used by \cite{lin/74} to indicate cases in 
which controllability was robust against perturbations of unknown or uncertain 
parameters. Here we use {\it structural}\, in a somewhat wider, but closely 
related, sense. All definitions of observability that classify a system in 
either observable or not will be included in the class of {\it structural 
observability}. The justification for this is that in such cases, observability 
only depends on the internal structure (presence and nature of coupling terms) 
of the system variables. Hence, 
in this sense, Kalman's definition of observability and the nonlinear 
counterpart \cite{her_kre/77} belong to this class although such are sometimes 
referred to as being definitions of {\it complete}\, or {\it full}\, observability. Other terms such as
{\it exact}\, and {\it mathematical}\, controllability/observability have been 
used recently \cite{wan_eal/17}.

A slightly different aim has been pursued in \cite{liu_eal/13} where a minimum 
set of sensor vertices is sought in order to render a graph observable. 
Nonetheless, the procedure either indicates situations in which the graph is or 
is not observable.

In spite of the varied terminology, there is one aspect common to all such 
procedures: the outcome is a classification of a system according to which it 
is either observable or not.  In view of this, we classify all such procedures 
under the heading of {\it structural observability}.  This is the case for the 
methods reviewed in sections~\ref{eon}, \ref{lm} and~\ref{lcm}.

\subsection{Dynamical Observability}
\label{dyo}

In contrast to structural observability, we shall refer to {\it 
dynamical observability}\, whenever there is a continuous 
quantification of our ability to estimate the state of a system from a finite 
set of data. In dynamical observability the key issue is to somehow quantify 
and distinguish those cases in which a system is close to becoming non 
observable from those in which it is far from that condition. This is done 
computing observability coefficients that measure how far from singularity, on 
average along a trajectory in state space, is the observability matrix, see discussion in 
sections~\ref{ros} and~\ref{slo}. Therefore, an important aspect of this class 
of observability is that it only makes sense for systems that {\it are}\, 
observable. Of course if a system is not observable, by definition the
corresponding dynamical observability coefficient is zero.  Hence dynamical 
observability helps us to rank {\it observable}\, pairs $[\mb{f},\,h_i(\bx)]$ 
for a given vector field $\mb{f}$.

A similar situation in terms of controllability of linear complex networks has 
been reported, namely the situation in which a network is controllable however, 
in practice, control is very difficult to attain \cite{wan_eal/17}. As argued 
by Cowan and coworkers: ``more important than issues of structural 
controllability are the questions of whether a system is almost 
uncontrollable'' \cite{cow_eal/12}. This is the typical situation in which a 
{\it dynamical}\, rather than a {\it structural}\, assessment of 
controllability or observability is called for. Dynamical observability was
investigated in the context of three-node networks of Fitzhugh-Nagumo 
oscillators in \cite{wha_eal/15}.

In assessing this type of observability, there are two challenges to be faced. 
First is how to quantify how far the system is, at a certain point, from the 
location in space where observability is lost, that is, where observability 
matrix becomes rank deficient. Second, how to average this result in order to 
have a single ``global'' indication of observability. In Sec.\,\ref{ros} these 
challenges were met by computing the condition number (\ref{do2}), and taking 
an average along a
trajectory (\ref{do3}) which can be interpreted as a spatial average in state space. If the trajectory 
happens to be chaotic, then the average covers a wider region than in the case of a periodic
trajectory. 

Other ways of facing the first challenge would be to use the determinant of the observability
matrix or its singular values. The fraction of time that the trajectory spends within a neighborhood 
of the singularity manifold has been used to assess dynamical observability \cite{fru_eal/12}. 

The coefficients that quantify dynamical observability have only relative 
interpretation. For instance, in the case of the R\"ossler system for 
$(a,\,b,\,c)=(0.398,\,2,\,4)$ the coefficients are ordered thus 
$\delta_y>\delta_x>\delta_z$ therefore the $y$ variable conveys better 
observability of the system than variable $x$ which, in turn, is preferrable 
to $z$.  Unfortunately, the coefficients for dynamical observability as
defined by the condition number (\ref{do2}), are not comparable in general 
among different systems. This shortcoming is overcome by the coefficients for 
symbolic observability, as discussed in Sec.\,\ref{so}.

\subsection{Symbolic Observability}
\label{syo}

{\it Symbolic observability}\, shares some features of the previous types of 
observability and includes characteristics of its own. On the one hand, as with 
structural observability, symbolic observability does not depend on parameter 
values but only on the nonlinear couplings within the system variables. On the 
other hand, as with dynamical observability, symbolic observability is
capable of ranking observable pairs.

Central to the definition of symbolic observability is the complexity of the 
singularities that appear in the symbolic observability matrix. Some advantages 
compared to the other definitions are the fact that it is more amenable to be 
computed for larger systems with more complicated dynamics \cite{bia_eal/15}, 
it provides ``normalized'' results in the range $[0; 1]$ that permit comparing 
different systems in terms of observability. Related to this, it has been 
argued that systems with a symbolic observability coefficient greater than 0.75 
have good overal observability properties \cite{sen_eal/16}.

These symbolic coefficients are very promising for 
assessing the observability of systems and networks that are larger than
the ones analyzed with the dynamical observability coefficients
\cite{let_eal/18}.

\section{Observability of Dynamical Networks: Numerical Results}
\label{net}

A dynamical network is a set of dynamical systems -- oscillators -- interconnected 
according to the network topology which is described by the corresponding adjacency matrix. 
The aim is to discuss, in the context of a simple example where the node dynamics is linear, 
some of the aspects seen so far.

Here we will consider a network whose topology is described by the adjacency
matrix
\begin{eqnarray}
  \label{adjac}
  A_{\rm adj} =
  \left[
  \begin{array}{ccc}
    0 & a_{12} & 0 \\
    a_{21} & 0 & a_{23} \\
    0 & a_{32} & 0
  \end{array}
  \right], 
\end{eqnarray} 

\noindent
and for which at each node there is a three-dimensional dynamical system
\begin{equation}
\label{lins3}
  S_i : 
  \left\{
    \begin{array}{l}
          \dot{x}_i = -y_i \\
      \dot{y}_i = x_i \\
      \dot{z}_i = \alpha x_i-z_i \, .
    \end{array}
  \right. 
\end{equation}

\noindent
Nodes are coupled via one of their variables ($x_i$, $y_i$ or $z_i$). In this
network, the term $a_{32}$ may vanish, for instance due to a nonlinearity. 
In what follows, we adopt the convention that the element $a_{ij}$ of the adjacency matrix 
$A_{\rm adj}$ corresponds to an edge from vertex $j$ to vertex $i$ \cite[Sec.\,6.2]{new/10}.
If the other convention were adopted, we would have to use $A_{\rm adj}^{\T}$ in place of
matrix $A$ or the Jacobian matrix. 

\begin{figure}[ht]
  \begin{center}
    \begin{tabular}{cc}
      \includegraphics[width=0.3\textwidth]{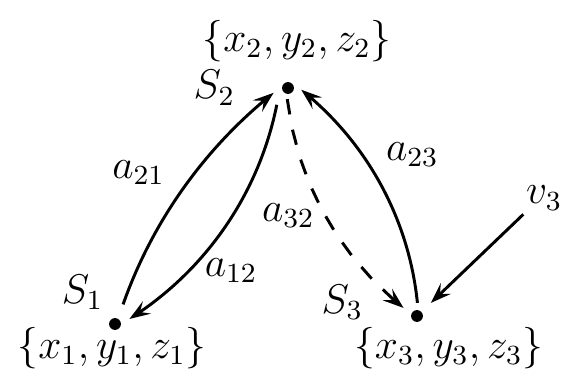} &
      \includegraphics[width=0.3\textwidth]{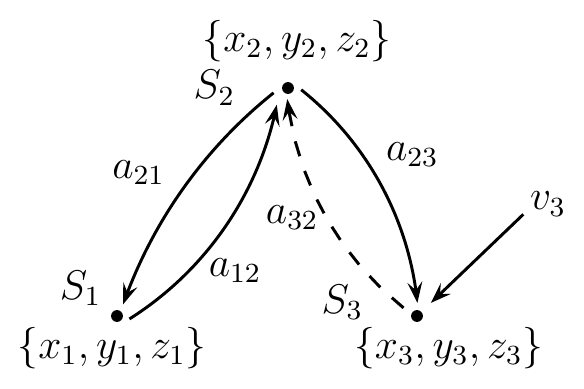} \\
      (a) Graph of the pair $[A_{\rm adj}, \mb{b}_3]$ &
      (b) Graph of the pair $[A_{\rm adj}^{\T}, \mb{c}_3]$ \\[-0.2cm]
    \end{tabular}
    \caption{\label{f070917b}(a)~Graph of the network whose topology is described
by the adjacency matrix (\ref{adjac}) with a single driving node $S_3$, and 
(b)~the corresponding dual graph.  If $a_{32}$ vanishes in (b), nodes 
$S_1$ and $S_2$ become non-accessible and the dual network is no longer 
structurally controllable, meaning that the pair $[A_{\rm adj}, \mb{c}_3^{\T}]$ is no 
longer structurally observable from $S_3$ if $a_{32}$ vanishes.}
  \end{center}
\end{figure}

Consequently, following
Newman's convention,  controllability can be investigated by considering the
pair $[A_{\rm adj},\,\bb]$ where the input vector is $\bb=[0~0~1]^{\T}$, 
indicating the situation in which only system $S_3$ receives the driving 
signal (Figure~\ref{f070917b}).  As long as $a_{32}\neq 0$ the network is structurally {\it 
topologically}\, controllable since each node can be reached from vertex $v_3$ 
(Figure~\ref{f070917b}a). The {\it topological}\, observability of the network 
can be analyzed using the dual pair $[A_{\rm adj}^{\T}, \bc]$ where the 
output 
vector is $\bc=[0~0~1]^{\T}$, indicating the situation in which only one or 
more variables from system $S_3$ can be measured (Figure~\ref{f070917b}b). As 
long as $a_{32}\neq 0$ the network is structurally topologically observable. A 
similar conclusion can be drawn directly from the graph shown 
Figure~\ref{f070917b}a but reversing the edge
from vertex $v_3$: as long as $a_{32}\neq 0$, information from nodes $S_1$ and 
$S_2$ can flow up to the measurements (vertex $v_3$) and, consequently, the 
network is structurally topologically observable.

Using symbolic observability and treating the adjacency matrix 
$A_{\rm adj}$ as a Jacobian matrix, it is readily found that the 
network in Figure~\ref{f070917b}a is not topologically observable from $S_2$ 
($\eta_2^3=0$), it is fully topologically observable from $S_1$ ($\eta_1^3=1$) 
and is poorly topologically observable from $S_3$ ($\eta_3^3=0.56$). The lack 
of observability from $S_2$, which can be readily
confirmed from linear system theory, is not obvious, as this node receives 
information from the other two nodes. This result seems to be in line with the 
discussion presented in \cite{lei_eal/17}.

\begin{figure}[ht]
  \begin{center}
    \begin{tabular}{ccc}
      \includegraphics[width=0.42\textwidth]{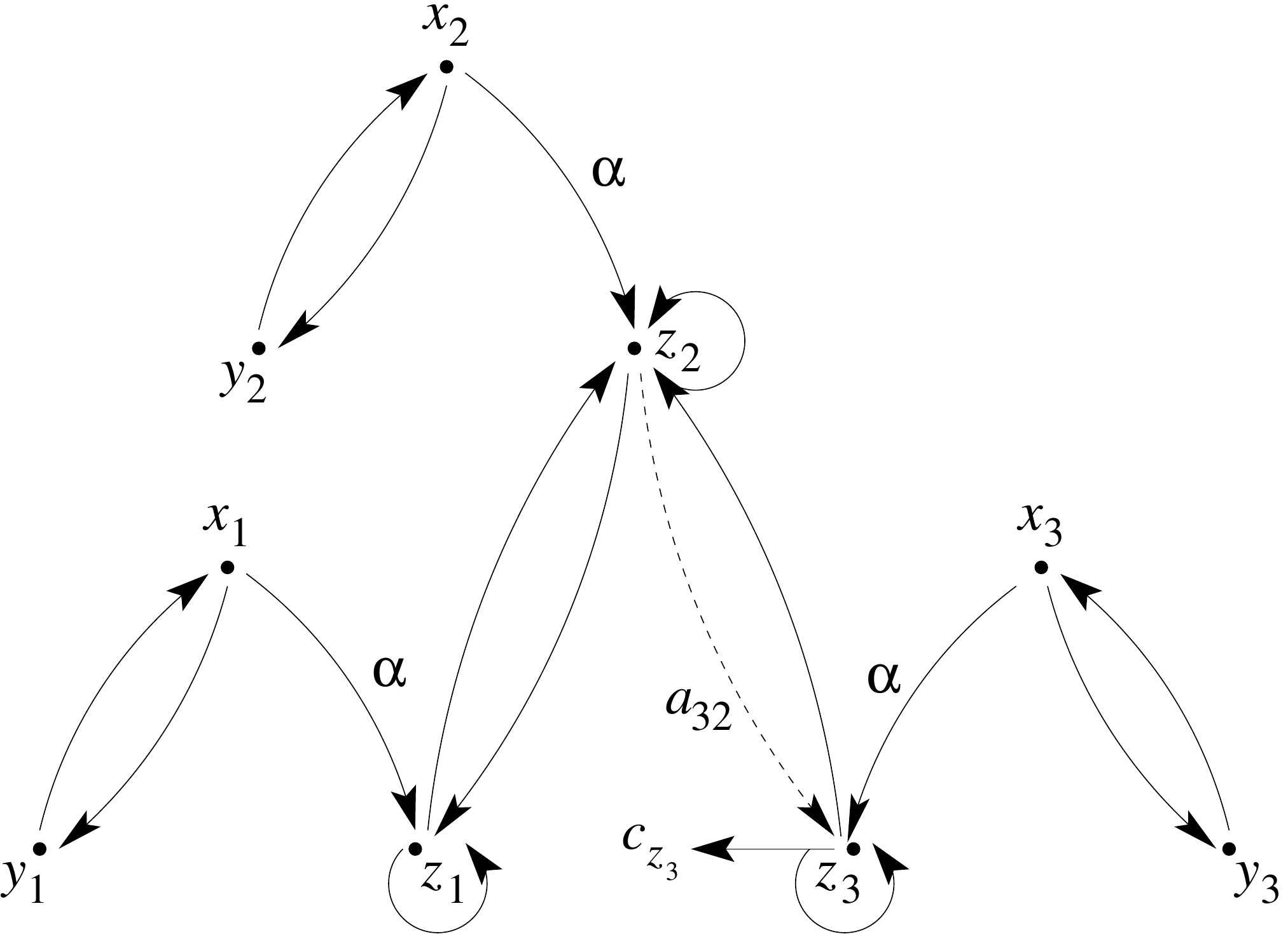} & ~~ &
      \includegraphics[width=0.42\textwidth]{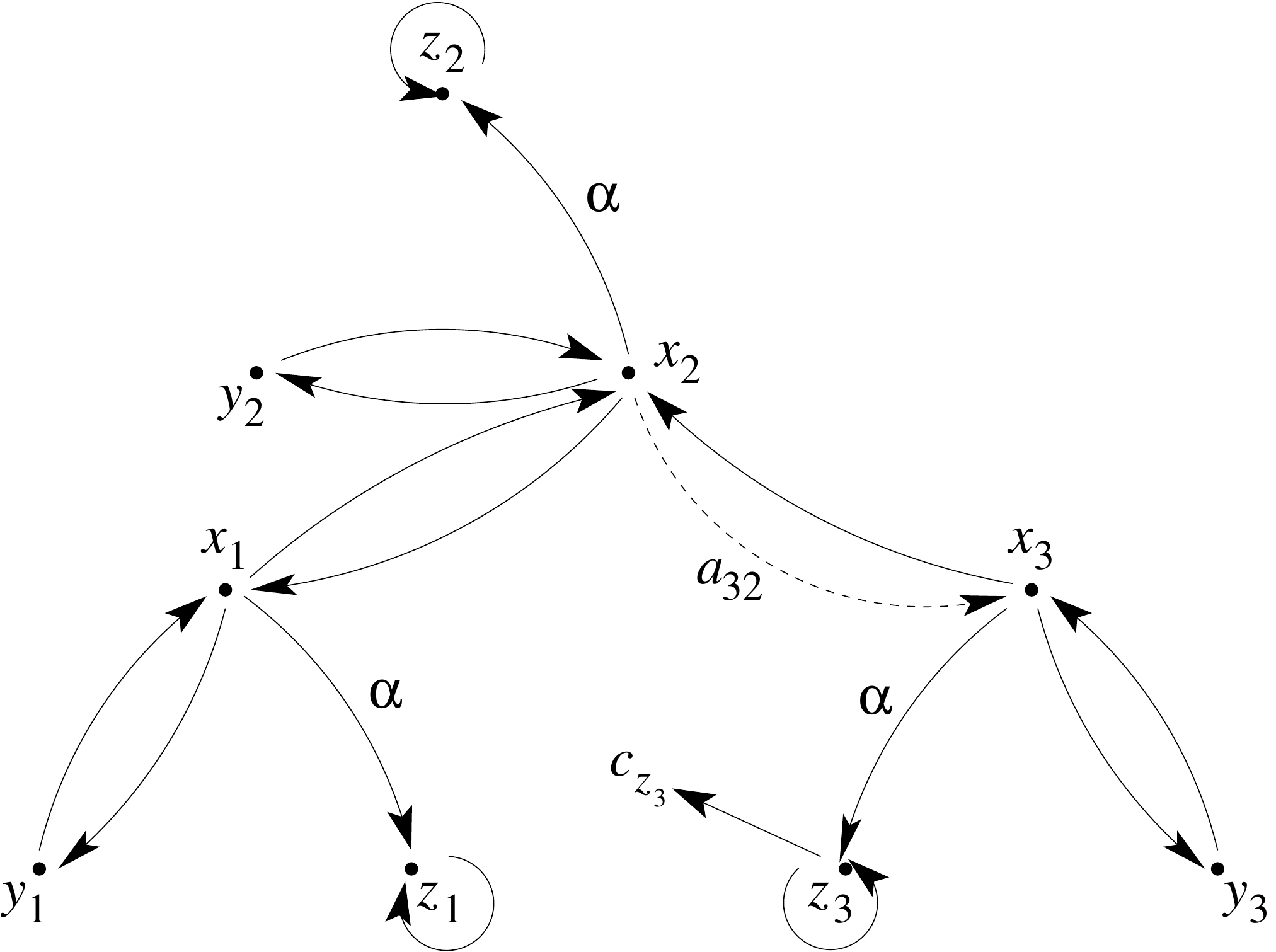} \\
      {\small (a) Coupled via variable $z$} & &
      {\small (b) Coupled via variable $x$} \\[-0.2cm]
    \end{tabular}
    \caption{Graph of the full network in Figure~\ref{f070917b}a. The details 
of the node dynamics are included. If $z_3$ is measured,  and $\alpha =0$ the 
network is not structurally observable regardless of the value of $a_{32}$. }
    \label{f070917c}  
      \end{center}
\end{figure}

Nevertheless, when considering the observability of a dynamical network as 
shown in Figure~\ref{f070917b}a with nodal dynamics (e.g. as given in Eq.\,\ref{lins3}), 
it must be realized that the topological 
observability only provides a partial answer. In order to ensure structural
observability of the full network from, say, $v_3$, not only every node of the dual
pair $[A_{\rm adj}^{\T}, [0~0~1]^{\T}]$ (Figure~\ref{f070917b}b) that must be 
accessible by acting on $v_3$ but also every vertex of the graph describing the full network as shown 
in Figure~\ref{f070917c}a. Consequently the 
result strongly depends on the observability conveyed by the variable  
used in measuring the sensor node {\it and}\, the one used for coupling the nodes. 

Since (\ref{lins3}) is linear, it is straightforward to verify that the 
pair $[D\mb{f}_3, \bc_3^{\T}]$ is structurally observable 
only if the measured variable is $z_3$ (that is, $\bc_3^{\T}=[0~0~1]$) and 
$\alpha \neq 0$. The pair $[D\mb{f}_3,\,\bc_3^{\T}]$ is not structurally observable when 
$\bc_3^{\T}=[1~0~0]$ or $\bc_3^{\T}=[0~1~0]$. Therefore, although the network is structurally 
{\it topologically}\, observable from $S_3$ ($a_{32}\neq 0$), it is 
only structurally observable if variable $z_3$ is recorded at $S_3$. In addition, if 
$\alpha =0$, the  network is {\it not}\, structurally observable even for 
$a_{32}\neq 0$. Indeed as shown by the unfolded graph of the network in 
Figure~\ref{f070917c}a, the node dynamics, $S_3$, is structurally observable when 
variable $z_3$ is measured ($\eta_{z_3^3} = 1$, 
Det~$\frac{\partial \Phi}{\partial \mb{x}} = - \alpha^2$ where $\Phi \, : 
(x_3, y_3, z_3) \mapsto (z_3, \dot{z}_3, \ddot{z}_3)$) and is not
observable when $x_3$ or $y_3$ ($\eta_{x_3^3} = \eta_{y_3^3} = 0$) are 
measured. 

When the nodes of the full network are coupled by variable $z$ there is a directed 
path from every vertex to vertex $z_3$ if $\alpha\neq 0$. To see that the 
network observability also depends on the coupling consider when 
coupling is accomplished via variable $x$ (or similarly via variable $y$). The unfolded graph 
drawn in Figure~\ref{f070917c}b shows that the resulting network will only be 
structurally observable if $z_1$, $z_2$ and $z_3$ are simultaneously recorded, 
even for $\alpha \neq0$ and $a_{32}\neq 0$.

The previous examples helps to understand why Gates and Rocha have 
argued that to represent nodes as variables lacks intrinsic dynamics and that 
there is often a discrepancy between results related to controllability that 
take into account only the graph structure \cite{gat_roc/16}.

To summarize, in investigating the observability of a dynamical network, 
not only the topology but also the local dynamics of sensor nodes and the 
coupling variables must be taken into account. For the sake of clarity, when a 
full network is investigated, these three ingredients must be considered: i)
nodes connected according to an adjacency matrix (a graph), ii)~the coupling 
 and iii)~the node dynamics.

\begin{figure}[ht]
  \begin{center}
    \includegraphics[scale=0.5]{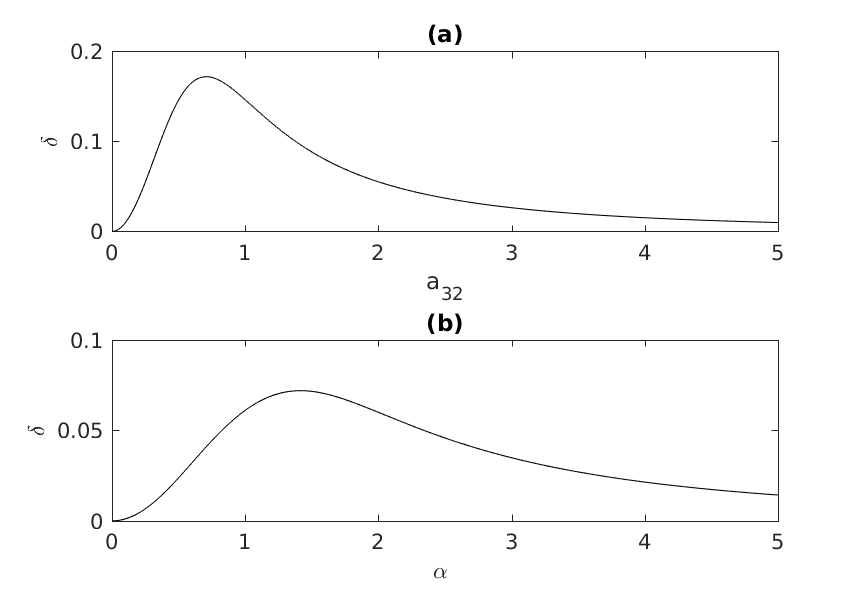} 
    \begin{tabular}{cc}
      \includegraphics[width=0.42\textwidth]{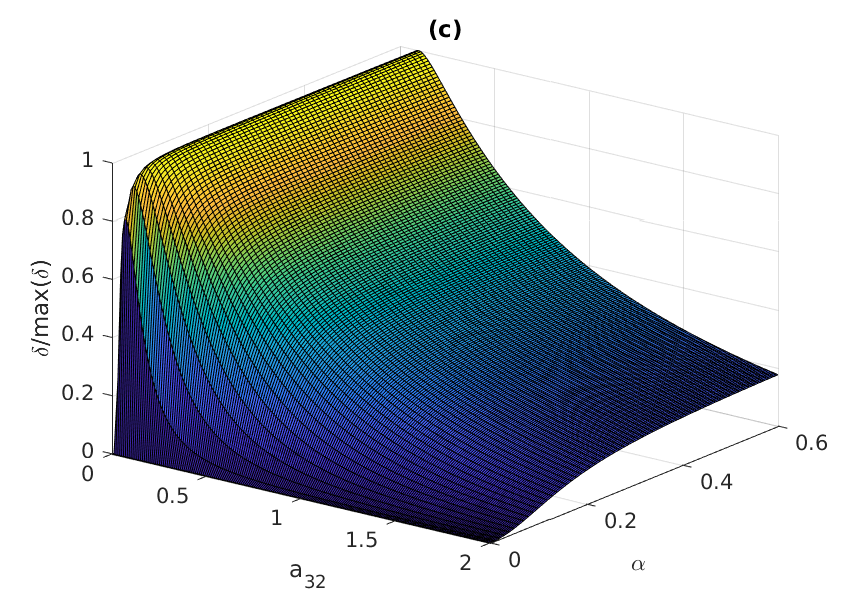} & 
      \includegraphics[width=0.42\textwidth]{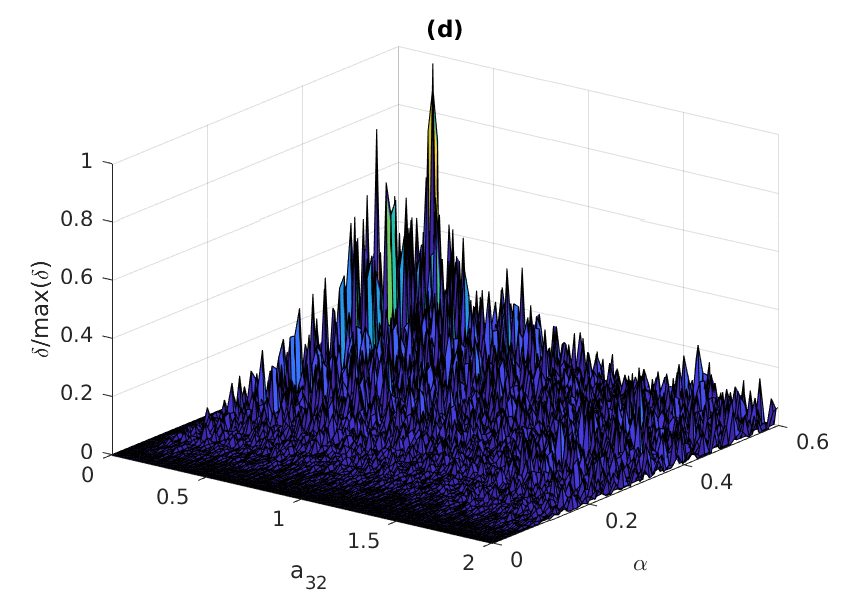} \\
    \end{tabular}
    \caption{\label{deltadelta}Observability coefficient (\ref{do}) computed 
(a)~for the network topology shown in Figure~\ref{f070917b}a, (b)~for 
the node dynamics in (\ref{lins3}). Mathematically, the node 
dynamics becomes not structurally observable from $z_3$ only for $\alpha=0$, 
and the network becomes structurally not topologically observable from the 
sensor node $S_3$ only for $a_{32}=0$. Normalized observability coefficients
(c)~for the full network coupled using $z$
(Fig.\,\ref{f070917c}a), and (d)~coupled using $x$ (Fig.\,\ref{f070917c}b). }
  \end{center}
\end{figure}

Structural topological observability of the full network (Figure~\ref{f070917c}) whose topology is 
described by $A_{\rm adj}$ (Figure~\ref{f070917b}a) will only detect when observability is 
completely lost and is insensitive to a gradual reduction in observability due 
to the decrease of $a_{32}$ or in $\alpha$. Here the simple procedure discussed 
in \cite{agu_ieee/94} will suffice to provide an indication of the gradual 
reduction in topological observability using (\ref{do}) applied to the 
adjacency matrix $A_{\rm adj}$ from Eq. (\ref{adjac}) if the dynamics are 
linear and,  using (\ref{do3}) if nonlinear. Dynamical observability of the 
toplogy of the network, disregarding the node dynamics, is shown in 
Figure~\ref{deltadelta}a whereas the dynamical observability of uncoupled node 
dynamics is shown in Figure~\ref{deltadelta}b. These plots resemble the overall 
shape of the plots presented in Ref.\,\cite[see their Fig.~5]{wha_eal/15} 
where for small coupling the observability coefficient is very low; it 
increases with the coupling up to a maximum, and then gradually falls as the 
coupling is further increased. 

For the full network in Figure~\ref{f070917c}a, 
that is, when three systems (\ref{lins3}) are coupled by their variable $z$ 
according to the adjacency matrix (\ref{adjac}) and when variable $z_3$ is 
measured the observability coefficients are shown in Figure~\ref{deltadelta}c.
Slices of this plot retains some features of the two previous ones. However, when the 
nodes are coupled through variable $x$ accordind to the {\it same}\, adjacency 
matrix, the observability is practically lost as illustrated in 
Figure~\ref{deltadelta}d, where the values in the plot are all close to zero within 
machine accuracy. This shows that a joint anlysis is required, that is, not 
only node dynamics and how the nodes are connected must be used, but also the 
coupling variables must be taken into account.

This result only provides a partial answer. A more acurate analysis can be
performed using Det$\frac{\partial \Phi_s}{\partial \mb{x}}$. 
First we set $a_{12}=a_{21}=0$ to treat a two node network $\{S_2, S_3\}$ 
with measurements in node $S_3$, either $x_3$, $y_3$ or $z_3$. We were able to get a full rank
$6 \times 6$ observability matrix with $\Phi_{x_3 y_3 z_3^4}$, 
$\Phi_{x_3^2 z_3^4}$ or $\Phi_{y_3^2 z_3^4}$ for which the determinants were
equal to $\pm a_{32}^3 \alpha^2$. The full network becomes structurally 
non observable when $\alpha_{32}$ or $\alpha$ is equal to zero as already
found. Therefore, the network is observable from $S_3$ if we measure $z_3$
plus, at least, another variable from that node.

This simple example shows that investigating a dynamical network by only 
analyzing the network observability from the adjacency matrix can lead to wrong 
results because the topological observability is only correct in the extreme
case where the nodes are not only coupled but also observed by the variable 
providing the best observability. 
If the network is structurally topologically observable, then the network 
observability depends on the variable with which node dynamics are coupled and 
observed. Consequently, topological observability must be at least associated
with an analysis of the observability of the node dynamics (the isolated system 
acting at each node) and it must be checked whether the coupling conveys the 
information up to the measured variable.

\section{Conclusions}
\label{conc}

Two decades have past since it was argued that a procedure borrowed from the 
theory of observability of linear systems could be adapted to explain why 
global modeling algorithms performed differently using different recorded 
variables \cite{let_eal/98}. This paper has aimed at providing a general view 
of how some concepts related to observability have developed in the realm of 
nonlinear dynamics and to point out some important differences among the approaches. In order to 
make distinctions clearer, some different types of observability measures were 
proposed. Also, the use of the discussed techniques in the field of dynamical 
networks has been discussed briefly. An overview is provided in 
Table~\ref{tab}.

An important point to realize is that whereas the definition of observability 
aims to classify a system as being observable or not, a more interesting 
challenge is to be able to rank variables {\it of observable systems}\, in 
terms of the potential performance each would have in certain practical 
situations. The first problem has been connected to structural observability, 
whereas the second one to dynamical and symbolic observability. 
These concepts can be readily applied to dynamical systems or to dynamical
networks and their three levels of description, namely: node dynamics, topology and the full network.

However, as for the observability of dynamical networks, some
limitations of graph-based procedures have been pointed out. It has been 
argued that the observability of a dynamical network depends on three 
ingredients: i)~the topology described by the adjacency matrix -- called 
topological observability in this paper--; ii)~the variable used for coupling 
nodes and iii)~the observability of node dynamics. It was shown that the 
topological observability of a network --- only based on the adjacency matrix 
--- can provide spurious assessment of the observability of the full network in 
certain cases. In the case of dynamical networks, which are composed of 
oscillators at the nodes interconnected according to a topology, topological  
observability does not seem adequate to accurately characterize a network 
dynamics.

{\footnotesize 

\begin{table}[htb]
  \centering
  \caption{\label{tab}Summary of types of observability and systems. Yes/No 
refers to practical applicability of numerical procedures discussed in the 
paper. The  ``Node dynamics'' corresponds to low-dimensional dynamical systems 
interconnected according to a ``Topology'' to form a dynamical ``Network''.}
  \begin{tabular}{ccccc}
    \hline \hline
    Type of &  Task & Node dynamics & Topology & Networks \\
    Observability  & & Sec.\,\ref{ods} &  Sec.\,\ref{gra} & Sec.\,\ref{net} 
    \\[0.1cm]    
    \hline 
    Structural &  observable vs. & Yes & Yes & Yes \\
      Sec.\,\ref{sto}         &   nonobservable & Sec.\,\ref{eon} & Sec.\,\ref{lm}--\ref{lcm}  \\ 
     & classification  &  &  \\[0.2cm]
    Symbolic &  Ranking variables & Yes & Yes & Yes \\
    Sec.\,\ref{syo}   &  & Sec.\,\ref{so} &  Sec.\,\ref{sog} \\[0.2cm]
    Dynamical & Ranking variables & Yes  & Only for small &   No \\
    Sec.\,\ref{dyo} &   &  Sec.\,\ref{ros}--\ref{ga} &dimension Sec.\,\ref{rog}   \\
    \hline \hline  \end{tabular} 
\end{table}

}

\section*{Acknowledgements}
L.A.A and L.P. gratefully acknowledge financial support by CNPq and CAPES.
The authors wish to thank Irene Sendi\~na-Nadal for stimulating discussions.

\bibliographystyle{apalike}


\end{document}